\documentclass[12pt,preprint]{aastex}

\usepackage{lscape}
\usepackage{subfigure}

\shorttitle{Radio structures of RLNLS1s}
\shortauthors{Gu et al.}

\begin{document}


\title{The radio properties of radio-loud narrow-line Seyfert 1 galaxies on parsec scales}


\author{Minfeng Gu\altaffilmark{1,2}, Yongjun Chen\altaffilmark{1,3}, S. Komossa\altaffilmark{4}, Weimin Yuan\altaffilmark{5}, Zhiqiang Shen\altaffilmark{1,3}, Kiyoaki Wajima\altaffilmark{6}, Hongyan Zhou\altaffilmark{7,8}, J. A. Zensus\altaffilmark{4}}





\altaffiltext{1}{Shanghai Astronomical Observatory, Chinese Academy of Sciences, Shanghai 200030, China;
gumf@shao.ac.cn}
\altaffiltext{2}{Key Laboratory for Research in Galaxies and Cosmology,
Shanghai Astronomical Observatory, Chinese Academy
of Sciences, 80 Nandan Road, Shanghai, 200030, China}
\altaffiltext{3}{Key Laboratory of Radio Astronomy, Chinese Academy of Sciences, 2 West Beijing Road, Nanjing, JiangSu 210008, China}
\altaffiltext{4}{Max-Planck-Institut f$\rm \ddot{u}$r Radioastronomie, Auf dem H$\rm \ddot{u}$gel 69, 53121 Bonn, Germany}
\altaffiltext{5}{Key Lab for Space Astronomy and Technology, National Astronomical Observatories, Chinese Academy of Sciences, Beijing 100012, PR China}
\altaffiltext{6}{Korea Astronomy and Space Science Institute, 776 Daedeokdae-ro, Yuseong, Daejeon 305-348, Korea}
\altaffiltext{7}{Polar Research Institute of China, 451 Jinqiao Road, Shanghai, 200136, China}
\altaffiltext{8}{Key Laboratory for Research in Galaxies and Cosmology, Department of Astronomy, University of Science and Technology of China,
Chinese Academy of Sciences, Hefei, Anhui, 230026, China}


\begin{abstract}
We present the detection of compact radio structures of fourteen radio-loud narrow line
Seyfert 1 (NLS1) galaxies from Very Long Baseline Array observations at 5 GHz, which
were performed in 2013. While 50\% of the sources of our sample show a compact core only,
the remaining 50\% exhibit a core-jet structure.
The measured brightness temperatures of the cores range from $10^{8.4}$ to $10^{11.4}$ K with a median value of $10^{10.1}$ K, indicating that the radio emission is from non-thermal jets, and that, likely, most sources are not strongly beamed, then implying a low
jet speed in these radio-loud NLS1 galaxies.
In combination with archival data taken at multiple frequencies,
we find that seven sources show flat or even inverted radio spectra, while steep
spectra are revealed in the remaining seven objects. Although all these sources
are very radio-loud with $R > 100$, their jet properties are diverse, in
terms of their milli-arcsecond (mas) scale (pc scale) morphology and their overall radio spectral
shape. The evidence for slow jet speeds (i.e., less relativistic jets), in combination
with the low kinetic/radio power, may offer an explanation for the compact VLBA
radio structure in most sources.
The mildly relativistic jets in these high accretion rate systems are consistent with a scenario,
where jets are accelerated from the hot corona above the
disk by the magnetic field and the radiation force of the accretion disk.
Alternatively, a low jet bulk velocity can be explained by low spin
in the Blandford-Znajek mechanism.

\end{abstract}





\section{Introduction}

While both permitted and forbidden optical emission lines are
present in narrow-line Seyfert 1 galaxies (NLS1s), their broad Balmer
lines are narrower than those of normal broad-line Seyfert 1 galaxies
with  FWHM($\rm H\beta$) less than 2000 $\rm km
~s^{-1}$ in NLS1s \cite[e.g.,][]{ost85,goo89}. Moreover, NLS1s also exhibit 
other extreme observational properties, such as relatively weak
forbidden-line emission, i.e., $\rm [O III] 5007/H\beta < 3$, 
strong permitted optical/UV Fe II emission lines, steep soft X-ray spectra,
and rapid X-ray variability. NLS1s are often thought to be young
AGNs with relatively small black hole masses and high accretion
rates \cite[review by][]{kom08}.

Conventionally, NLS1s were thought to be radio quiet. With the 
discovery of radio loud NLS1s
(RLNLS1s), it has been realized, that NLS1s simply have a low probability 
to be radio loud, but are not completely radio quiet \citep{kom06,zho06}. 
Interestingly, RLNLS1s are inhomogeneous in their radio properties.
As shown in \cite{kom06}, most RLNLS1s in their sample are 
compact, steep spectrum sources, and hence likely associated with compact 
steep-spectrum (CSS) radio sources. 
A few NLS1s, however, showed flat or inverted radio spectra, and 
other properties characteristic of blazars. In fact, further observational
evidence has recently accumulated, that a significant fraction of
RLNLS1s, especially at the highest radio-loudnesses, does display
the characteristics of blazars, including large-amplitude 
radio flux and spectral variability, compact radio cores, 
very high variability brightness
temperatures, enhanced optical continuum emission, 
flat X-ray spectra, and blazar-like spectral
energy distributions (SED) \cite[e.g.,][]{yua08}, and several of them have been detected in 
the $\gamma-$ray regime with $Fermi$ for the first time \cite[][]{abd09a,abd09b}.

Very long baseline interferometry (VLBI) is one of the most
powerful tools for revealing the radio properties at parsec (pc)
scales by direct imaging at milli-arcsecond resolution. 
The early Japanese VLBI Network 
observations of a few RLNLS1s favored the presence of
relativistic jets, based on their high brightness temperatures, and their 
inverted radio spectra \citep{doi07}. 
Besides confirming the presence of relativistic jets, the previous 
work based on VLBA data argued that RLNLS1s can be either intrinsically radio loud,
or apparently radio loud due to jet beaming effects 
\citep{gu10,doi11}. However, due to the very limited numbers of RLNLS1s observed with VLBI \cite[e.g.][]{gu10,doi11,gir11,dam13,waj14,ric15}, 
it is far from clear that these features are universal in RLNLS1s.

Despite the rapidly growing number of RLNLS1s \citep{zho06}, 
the mechanisms driving their radio properties are still unclear, 
including the role of accretion rate, black hole spin, circum-nuclear matter, host galaxy and merger history. In addition to radio observations, the detection 
of flaring $\gamma$-ray emission has confirmed the presence of relativistic jets in a few
RLNLS1s \citep{abd09a,abd09b}. 
These observations may possibly be contrary to the well-known paradigm that
jets are generally associated with elliptical host galaxies in typical radio-loud AGNs, 
while there is tentative evidence suggesting that at least a few low-redshift RLNLS1s are hosted by spiral galaxies \cite[e.g.][]{zho07}. Further,
RLNLS1s generally have small black hole masses \cite[e.g.][]{yao15a}, high accretion rates, and very strong Fe II emission \cite[e.g.][]{kom06,yua08}, and 
are therefore at an opposite end of AGN correlation space than classical radio-loud AGNs \cite[e.g.,][]{sul08}. 

With their remarkable multi-wavelength properties and extreme location
in AGN parameter space, 
RLNLS1s allow us to re-address some of the key 
questions regarding the physics of jet formation, for example, the physical conditions under which a jet can be launched.
Understanding the extreme properties (e.g., steep and 
flat spectra) shown simultaneously 
in the same class of objects, would be of help in determining the 
physical conditions of the jets, and in the core regions of these galaxies. Understanding the
differences from other types of AGNs may then give further
important insights into jet formation \cite[e.g.,][]{kom15}. 
However, so far, little is known about the jets in RLNLS1,
primarily due to the lack of systematic observations 
of their jets with high angular resolution.

In this paper, we present the first systematic
high resolution radio study of RLNLS1s from the VLBA observations of
14 sources. The sample is presented in Section 2. The
observations and data reduction is shown in Section 3, and the
results are given in Section 4. Section 5 is for the discussion, and the last section is dedicated to
conclusions. Throughout this paper, we assume a cosmology with
$H_{0}=71~\rm km~s^{-1}~Mpc^{-1}$, $\Omega_{\rm M}=0.27$, and
$\Omega_{\Lambda}=0.73$ \citep{spe03}. The spectral index $\alpha$
is defined as $f_{\nu}\propto\nu^{\alpha}$, in which $f_{\nu}$ is
the flux density at frequency $\nu$.

\section{Sample}

To systematically study the radio structure at high resolution, we firstly compiled a large sample of RLNLS1s with the radio loudness parameter
$R=f_\nu \rm (1.4 GHz)$$/f_\nu \rm (4400{\AA})\ge10$, by combining various samples available so far.
The bulk of our sample is composed of the RLNLS1s of 
\cite{kom06}, \cite{zho06}, and \cite{yua08},
totaling 99 objects.
Furthermore, 18 sources from the compilation of \cite{fos11} are added in. 
The final sample thus consists of 117 RLNLS1s with $R\ge10$,
making it the  largest RLNLS1 sample so far. 
Among the sample, only eight sources were detected by $Fermi$ LAT \citep{abd09a,abd09b,dam12,dam15,fos11}.
We searched in the literature and found that, at that time, only 10 sources had published VLBI observations,
six with $Fermi$ detection and four without.
We excluded these ten sources from our VLBA target list, 
along with two in the southern hemisphere
 (RX J0134$-$4258 and PKS 0558$-$504). 
As a result, our final RLNLS1 sample without published VLBI observations
consists of 105 sources, all of which 
have 1.4 GHz flux densities measured from the Faint Images of the Radio Sky at Twenty
Centimeters (FIRST) 1.4-GHz radio catalogue \citep{bec95}. This is the first of a series of papers to systematically study the jet properties of RLNLS1s, aiming at gaining insights into the jet formation in these sources. 

\section{Observations and data reduction}

As the first step of our systematic studies on the whole sample, we have performed VLBA observations with a total observing time of 15 hours for 16 sources on October and November, 2013 (program ID: BG217), selected with 1.4 GHz flux densities $\ge$10 mJy from 105 objects without published VLBI observations at that time. The key scientific goal is to image the putative relativistic jets in these objects at high resolution. Results have been obtained for 14 sources (including one $Fermi$ object) with sufficient data quality to make images. 
The source list is shown in Table \ref{source}, in which the FIRST 1.4 GHz, the NRAO VLA Sky Survey
(NVSS) 1.4 GHz \citep{con98}, the 5 GHz flux density from the Green Bank 6-cm (GB6) survey (when available) \citep{gre96}, and the conventional radio loudness $R$ are given. 
All these sources are from \cite{yua08}, except for SDSS J142114.05+282452.8 from \cite{fos11}.

Observations were made with the VLBA at the C band ranging from 3.9 to  
7.9 GHz. We took the  observational mode RDBE/DDC to use the  
largest recording rate of 2 Gbps, corresponding to a recording bandwidth of 256 MHz in each of the two polarizations, to achieve sufficiently high imaging sensitivity.
Dual circular polarizations were recorded at two widely separated 
frequencies of about 5 GHz (4.868 GHz at 4.804 - 4.932 GHz) and 6.7 GHz (6.668 GHz at 6.608 - 6.736 GHz) so that the polarization and spectral information may
be obtained simultaneously. Sixteen target sources were observed in three 
separate blocks on Oct. 25, Nov. 14, and Nov. 24, 2013. The observations were made in phase referencing mode for three objects with the lowest flux densities (i.e. SDSS J085001.17+462600.5, SDSS J113824.54+365327.1 and SDSS J163401.94+480940.2, see Table \ref{source}). The on-source observing time was about 50 minutes for 
these three sources, while it was about 30 minutes for the remaining sources. Post data processing was performed within the package AIPS in 
a standard way for polarimetric VLBI observations, with the phase-referencing
method applied to the three weakest sources. Due to its extremely low linear
polarized emission, OQ 208 was successfully used to obtain the instrumental polarization 
in all the observations. As for the absolute polarized positional 
angle calibration, the calibrator 3C 286 was used but only a small part of the total flux density  
is restored, and hence only the fractional polarization images are given
in this paper. It should be noted that the scaling factors in amplitude calibration 
for the left circular polarization of several antennas are obviously incorrect,
therefore, we used the model of OQ 208 obtained from the right-handed polarization to calibrate the amplitude of the left-handed polarization, which proves to be 
 quite successful. Imaging and model-fitting were done with DIFMAP, and 
 the final results are given in Table \ref{modelfit}. The typical uncertainties in the flux density and size are about 
5\%, which is estimated by using the approximation in \cite{fom99}. 






\section{Results}

\subsection{Morphology and brightness temperature}

\begin{figure}
  \begin{center}
    \mbox{
     \subfigure
     {\scalebox{0.38}{\includegraphics{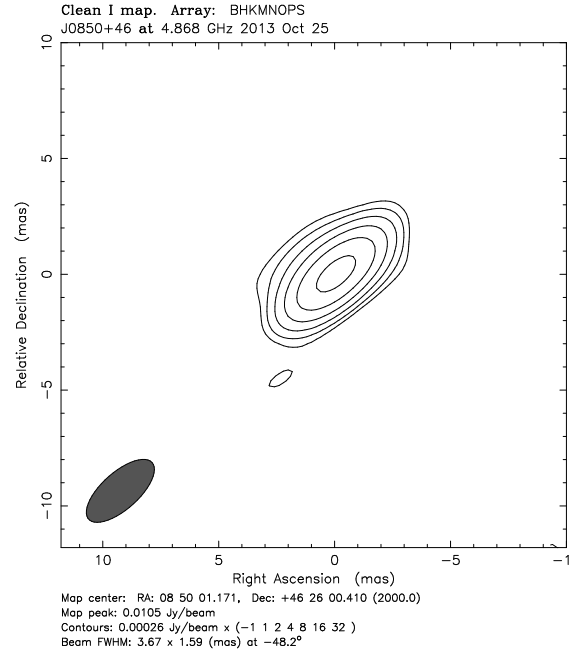}}}
      \quad

}

  \end{center}
  \caption{VLBA 5 GHz image of SDSS J085001.17+462600.5.} \label{j0850}
\end{figure}

\begin{figure}
  \begin{center}
    \mbox{
     \subfigure
     {\scalebox{0.4}{\includegraphics{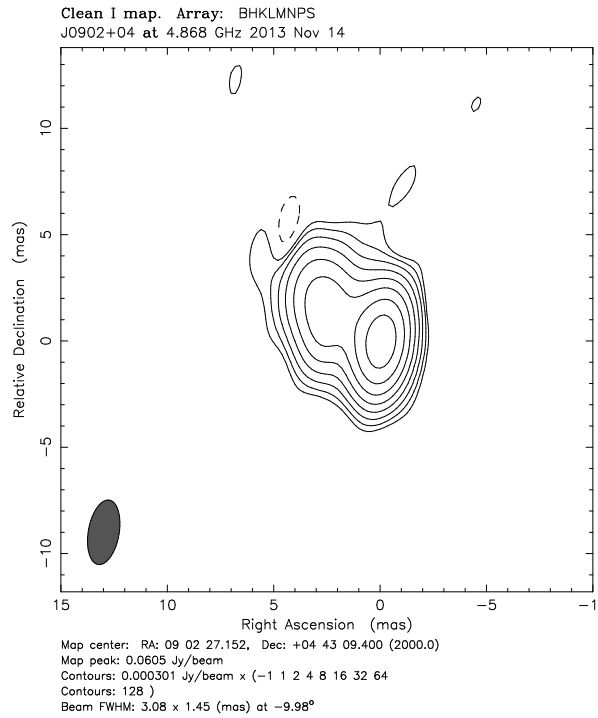}}}
      \quad

}

  \end{center}
  \caption{VLBA 5 GHz image of SDSS J090227.16+044309.6.} \label{j0902}
\end{figure}

\begin{figure}
  \begin{center}
    \mbox{
     \subfigure
     {\scalebox{0.38}{\includegraphics{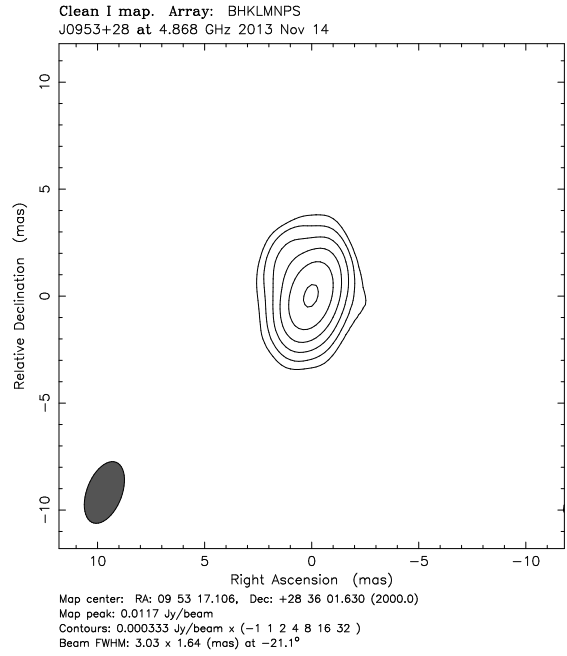}}}
      \quad

}

  \end{center}
  \caption{VLBA 5 GHz image of SDSS J095317.09+283601.5.} \label{j0953}
\end{figure}

\begin{figure}
  \begin{center}
    \mbox{
     \subfigure
     {\scalebox{0.4}{\includegraphics{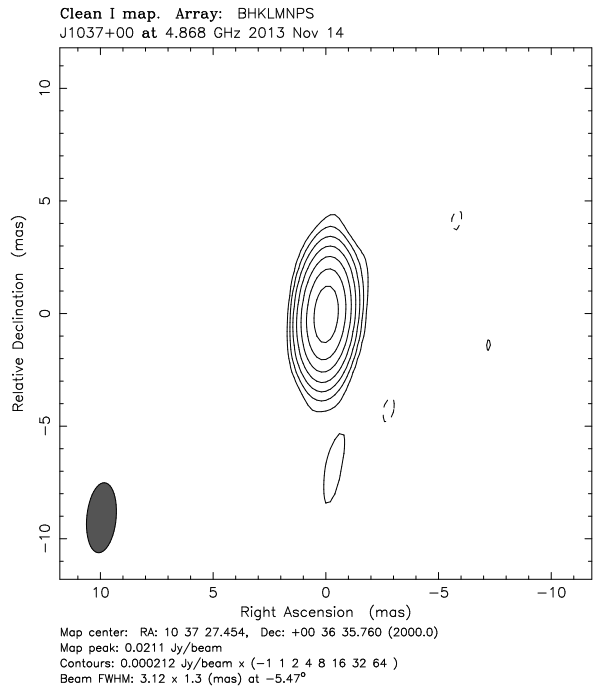}}}
      \quad

}

  \end{center}
  \caption{VLBA 5 GHz image of SDSS J103727.45+003635.6.} \label{j1037}
\end{figure}

\begin{figure}
  \begin{center}
    \mbox{
     \subfigure
     {\scalebox{0.38}{\includegraphics{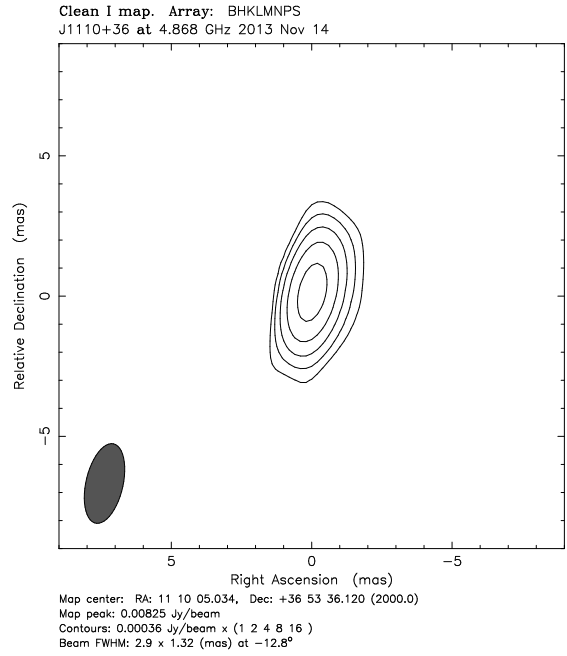}}}
      \quad

}

  \end{center}
  \caption{VLBA 5 GHz image of SDSS J111005.03+365336.3} \label{j1110}
\end{figure}

\begin{figure}
  \begin{center}
    \mbox{
     \subfigure
     {\scalebox{0.4}{\includegraphics{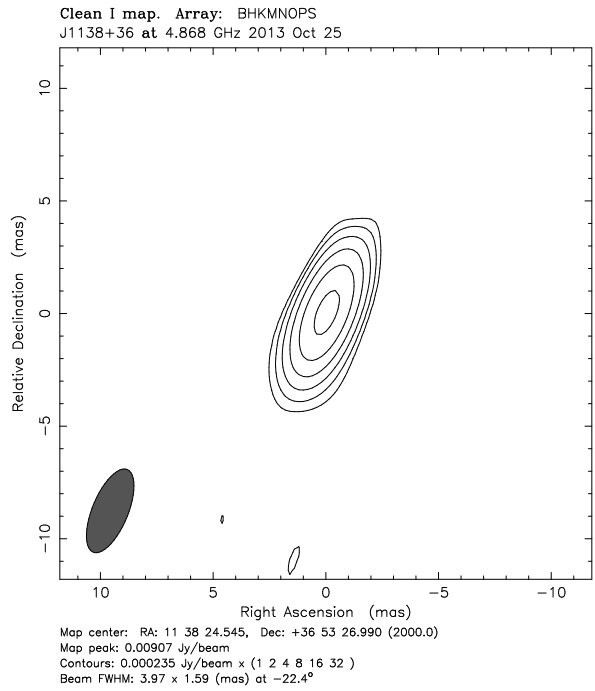}}}
      \quad

}

  \end{center}
  \caption{VLBA 5 GHz image of SDSS J113824.54+365327.1.} \label{j1138}
\end{figure}

\begin{figure}
  \begin{center}
    \mbox{
     \subfigure
     {\scalebox{0.4}{\includegraphics{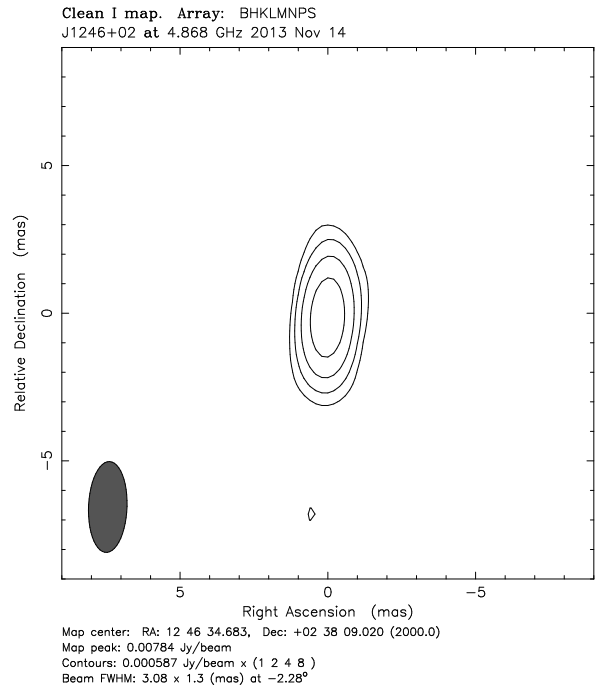}}}
      \quad

}

  \end{center}
  \caption{VLBA 5 GHz image of SDSS J124634.65+023809.0.} \label{j1246}
\end{figure}

\begin{figure}
  \begin{center}
    \mbox{
     \subfigure
     {\scalebox{0.4}{\includegraphics{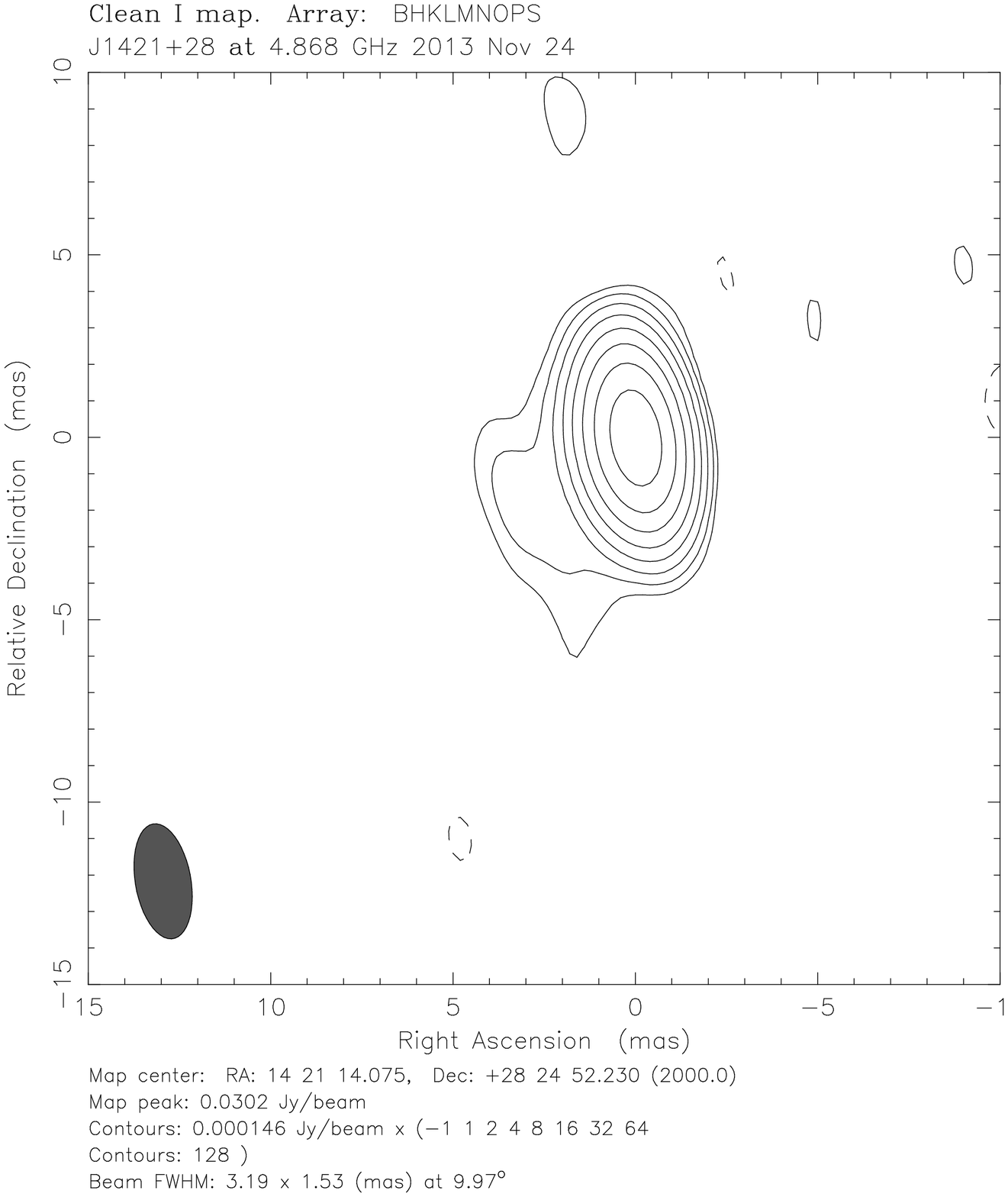}}}
      \quad

}

    \mbox{
     \subfigure
     {\scalebox{0.4}{\includegraphics{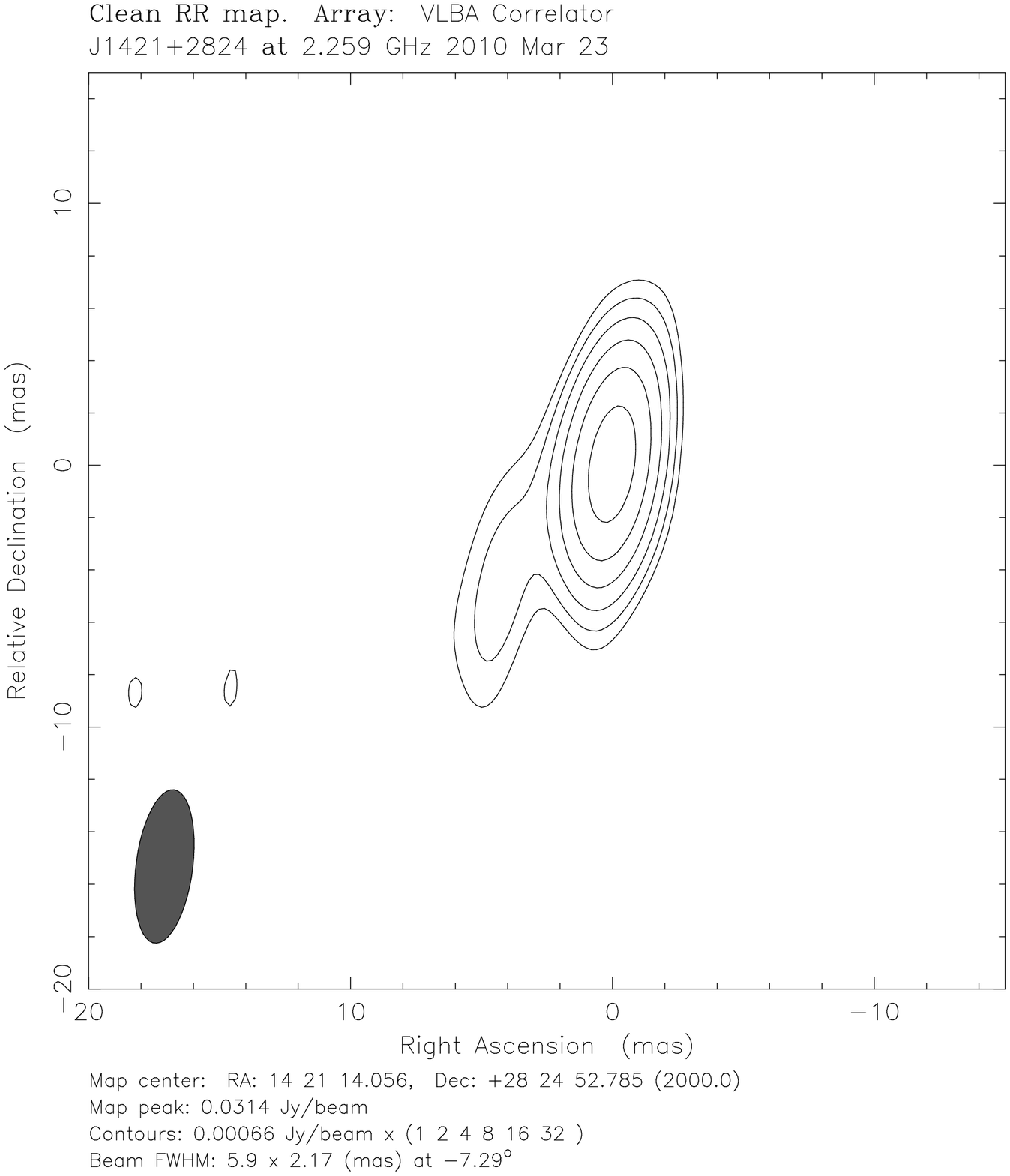}}}
      \quad

      \subfigure
      {\scalebox{0.4}{\includegraphics{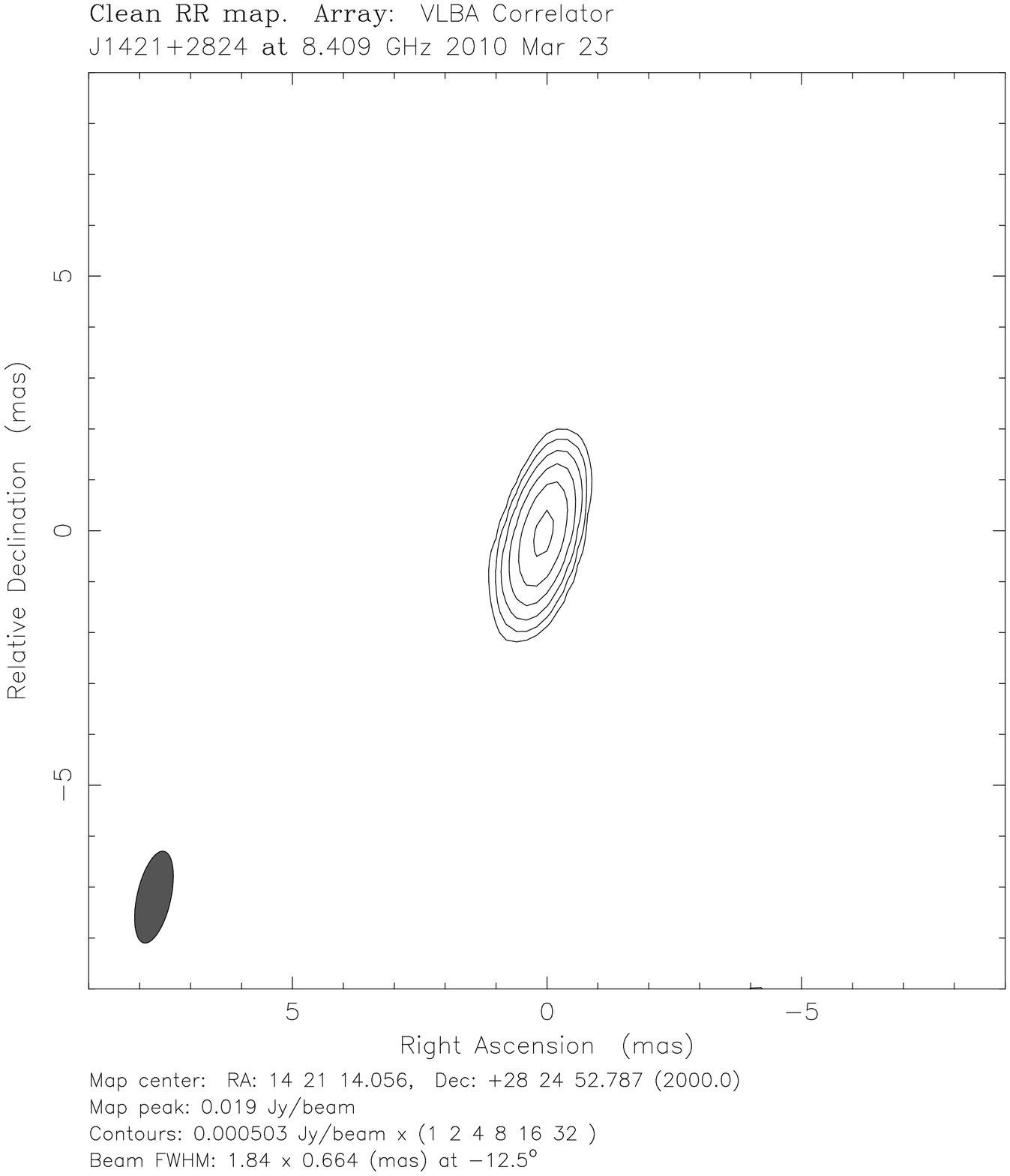}}}
      \quad}
  \end{center}
  \caption{VLBA 5 GHz image of SDSS J142114.05+282452.8, and archival VLBA 2.3 and 8.4 GHz images.} \label{j1421}
\end{figure}

\begin{figure}
  \begin{center}
    \mbox{
     \subfigure
     {\scalebox{0.4}{\includegraphics{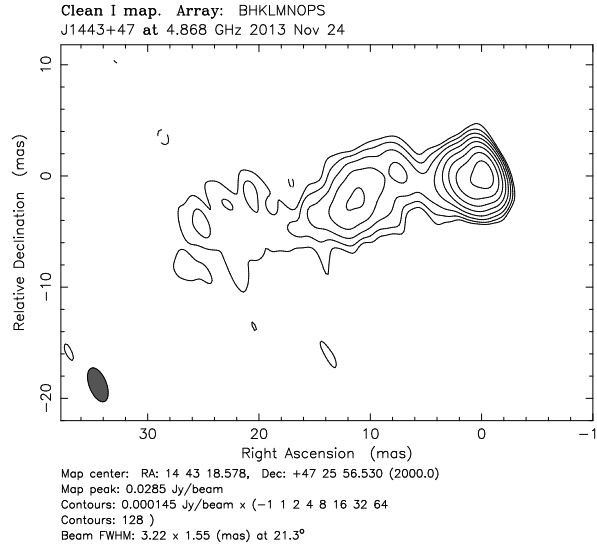}}}
      \quad

}

  \end{center}
  \caption{VLBA 5 GHz image of SDSS J144318.56+472556.7.} \label{j1443}
\end{figure}

\begin{figure}
  \begin{center}
    \mbox{
     \subfigure
     {\scalebox{0.38}{\includegraphics{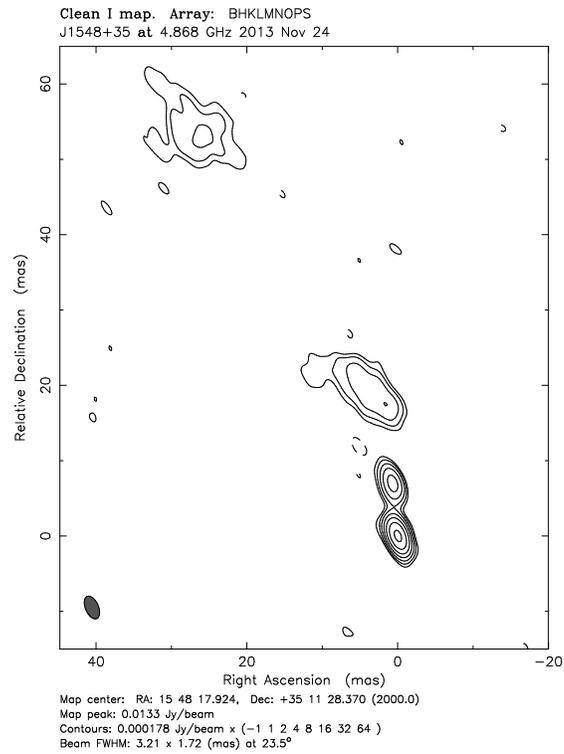}}}
      \quad

}

  \end{center}
  \caption{VLBA 5 GHz image of SDSS J154817.92+351128.0.} \label{j1548}
\end{figure}

\begin{figure}
  \begin{center}
    \mbox{
     \subfigure
     {\scalebox{0.38}{\includegraphics{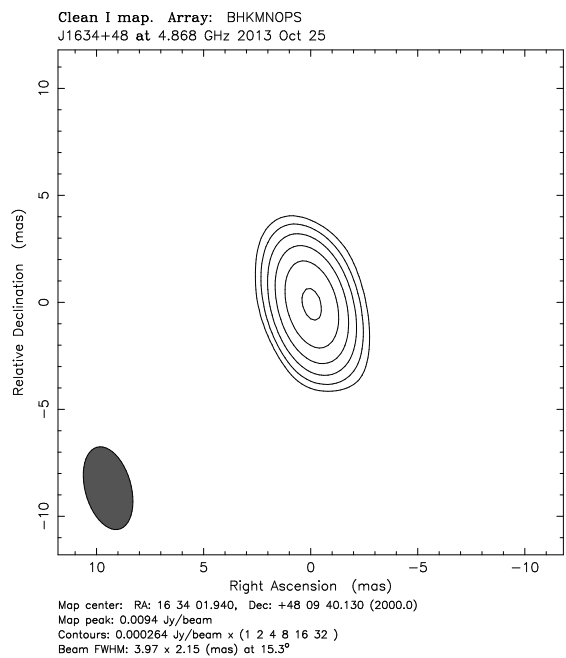}}}
      \quad

}

  \end{center}
  \caption{VLBA 5 GHz image of SDSS J163401.94+480940.2.} \label{j1634}
\end{figure}

The VLBA 5 GHz images for each object are shown in Figs. \ref{j0814} - \ref{j1634}. We find that seven sources show a compact core only, while the remaining seven objects have core-jet structure. When available, we also present the images from archival VLBA data for these sources. The radio core of each source is identified as a component with a flat spectrum of $\alpha>-0.5$ tentatively estimated with 5 and 6.7 GHz flux density from our measurements, and using archival VLBA multi-band data for several objects as well (see Section \ref{ind.obj.}). All
sources are compact with all components being within $\sim5$ mas ($\sim$ 24 - 38 pc depending on source redshift), except for two sources (SDSS J144318.56+472556.7 and SDSS J154817.92+351128.0) with extended jet emission at a distance of $>20$ mas (see Section \ref{ind.obj.}). Our results thus indicate that the majority of our sources is compact at VLBI mas scales.

From the high-resolution VLBA images, the brightness temperature of
the radio core $T_{\rm B}$ in the rest frame can be estimated with
\citep{ghi93}

\begin{equation} T_{\rm B}=\frac{S_{\nu}\lambda^2}{2k\Omega_{\rm
s}}=1.77\times10^{12}(1+z)(\frac{S_{\nu}}{\rm Jy})(\frac{\nu}{\rm
GHz})^{-2}(\frac{\theta_{d}}{\rm mas})^{-2} \end{equation} in which
$z$ is source redshift, $S_{\nu}$ is the core flux density at frequency
$\nu$, and $\theta_{\rm d}$ is the source angular diameter $\theta_{\rm
d}=\sqrt{\rm ab}$ where a and b are the major and minor axis,
respectively. This estimate actually represents a lower limit on the brightness temperature. 
The Doppler factor of the jet $\delta$ can be usually constrained by 

\begin{equation} T_{\rm B}^{'}=T_{\rm B}/\delta \label{tb} \end{equation} in which $T_{\rm B}^{'}$ is the intrinsic brightness temperature. Normally, $T_{\rm B}^{'}$ can be taken as
the equipartition brightness temperature of $5\times10^{10}$ K \citep{rea94}, or the inverse Compton catastrophe brightness temperature of $\sim10^{12}$ K \citep{kel69}. We note in passing that the Doppler factor estimated from radio observations is generally smaller than that from the SED modelling, specifically for $\gamma-$ray sources \cite[e.g.,][]{abd09b}. This could be due to the fact that $\gamma-$rays are produced in very compact regions. In contrast, the radio emission is likely from much larger regions, in which the jet blob is expanded to be optically thin, therefore, it likely has decelerated \cite[e.g.,][]{bla79}.

\begin{figure}
\epsscale{1.}
\plotone{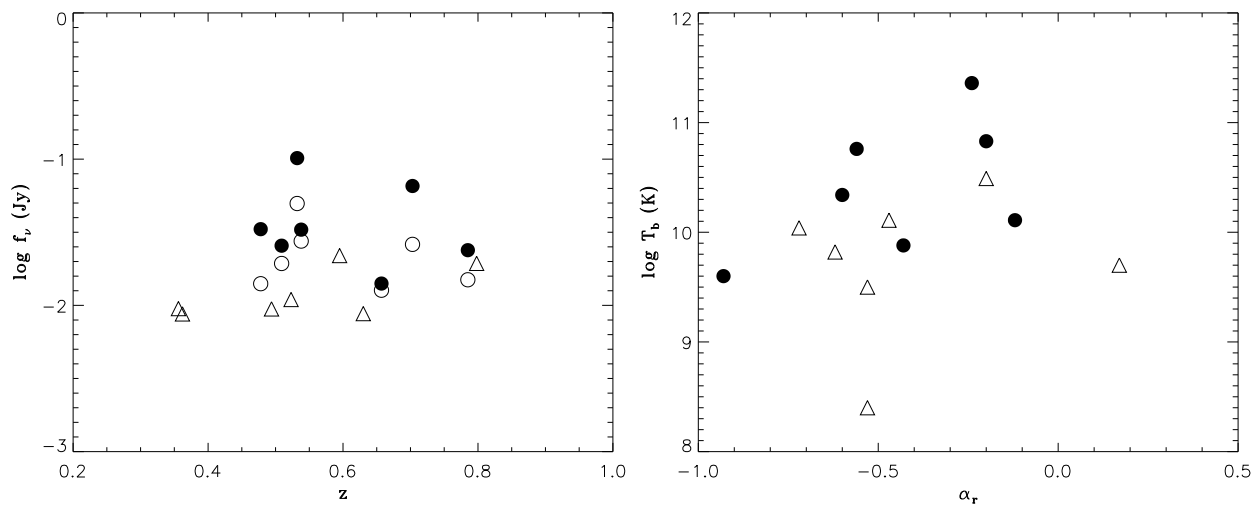}
\caption{Left: Source redshift versus flux density. Triangles are for objects with core only (and then the flux densities are for the cores). 
The circles are sources with core-jet morphology. The solid circles represent the overall source flux density, while the open circles are
for the core only. Right: Core brightness temperature versus radio spectral index based on low-resolution NED data (see text for details). The circles are for core-jet sources, while the triangles represent
core-only objects.\label{ccj}}
\end{figure}

The core brightness temperature ranges from $10^{8.4}$ to $10^{11.4}$ K with a median value of $10^{10.1}$ K (see Table \ref{modelfit}). This indicates that the radio emission is from non-thermal jets, confirming that powerful jets can be formed in accretion systems with relatively small black hole masses and high accretion rates \cite[see e.g.][]{gu10,doi11}. In contrast, the VLBA core brightness temperatures of blazars typically range between $10^{11}$ and $10^{13}$ K with a median value near $10^{12}$ K, and can even extend up to $5\times10^{13}$ K \citep{kov05,kov09}. While these measurements represent lower limits, the brightness temperatures of most our sources are below the equipartition and inverse Compton catastrophic limits. Therefore, the beaming effect is generally not significant in our sources. In principle, the strong Doppler beaming requires not
only a nearly pole-on view but also a relativistic bulk jet speed. Thus, the lack of a strong beaming effect implies that the bulk jet speed may likely be low in our sources, which still needs to be confirmed with higher-resolution observations, since measured brightness temperatures are lower limits. Alternatively, it is possible that our sources are seen at slightly larger viewing angles. We find that the RLNLS1s with core only morphology generally have lower core and total flux density, lower redshift and lower brightness temperature than those of core-jet objects (see Fig. \ref{ccj}). This implies that the former likely have less powerful jets and are less beamed compared to the latter.

\subsection{Polarization}

\begin{figure}
  \begin{flushleft}
    \mbox{
     \subfigure
     {\scalebox{0.37}{\includegraphics{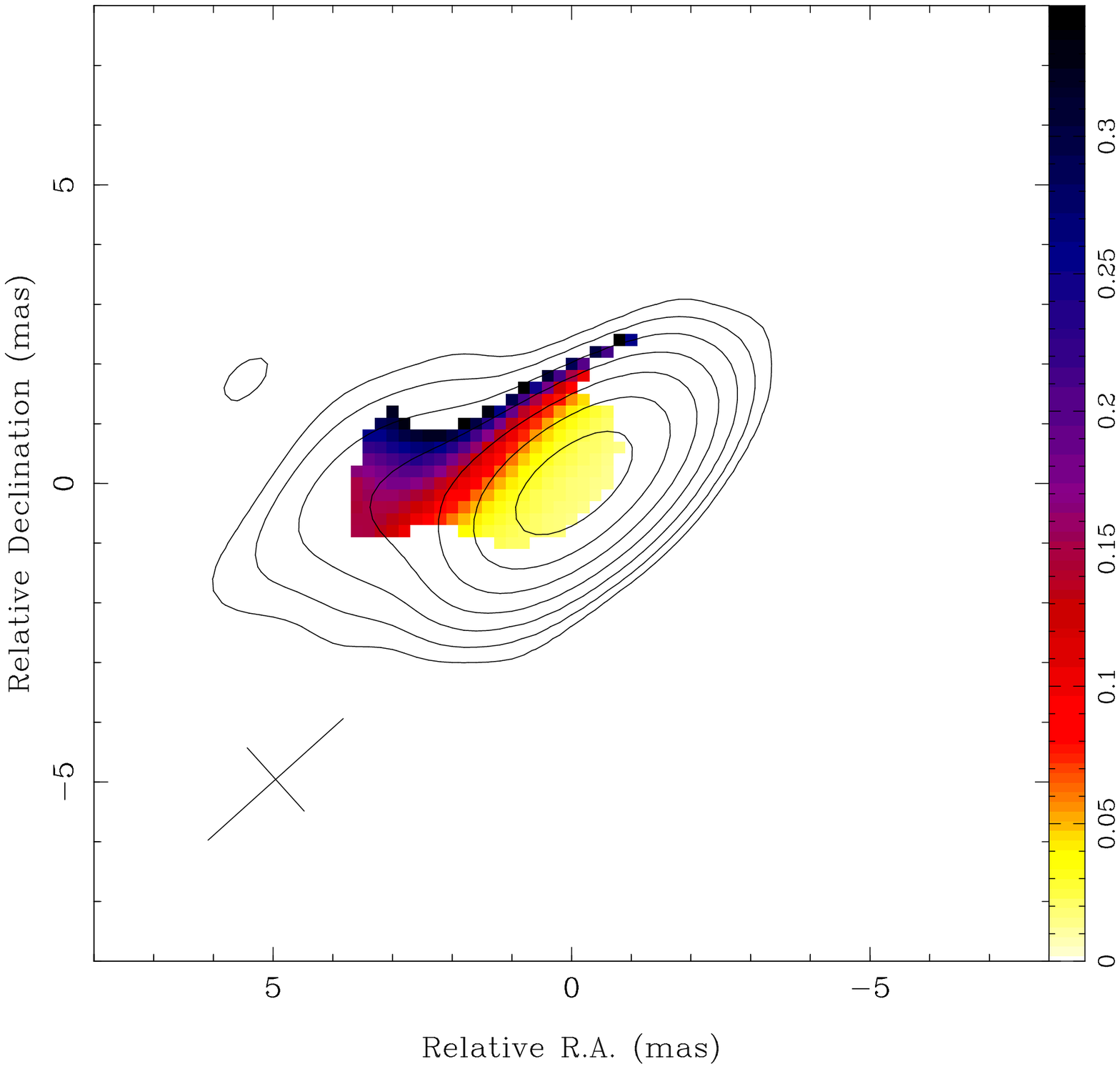}}}
      \quad

      \subfigure
      {\scalebox{0.36}{\includegraphics{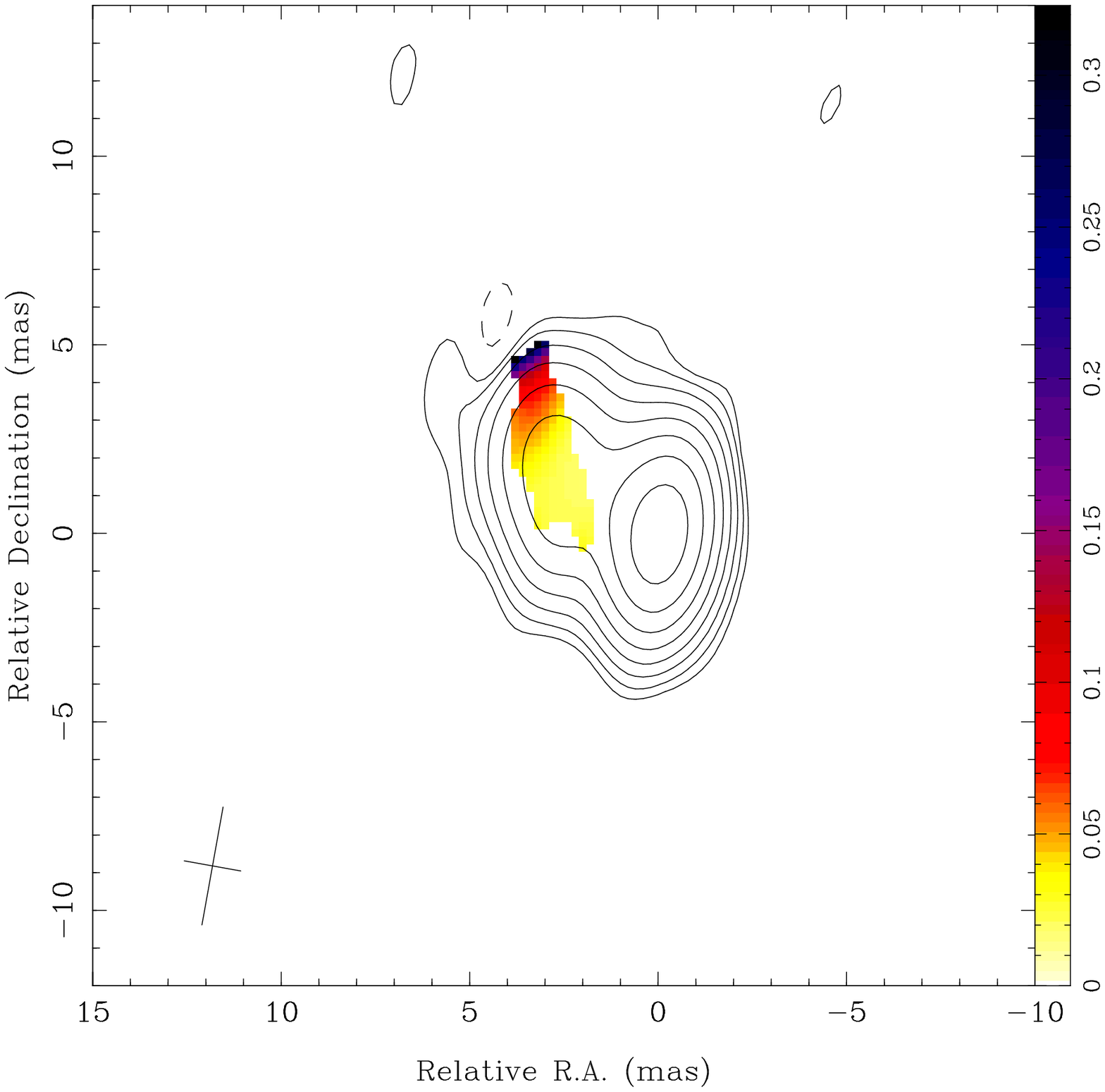}}}
      \quad}

    \mbox{
      \subfigure
      {\scalebox{0.37}{\includegraphics{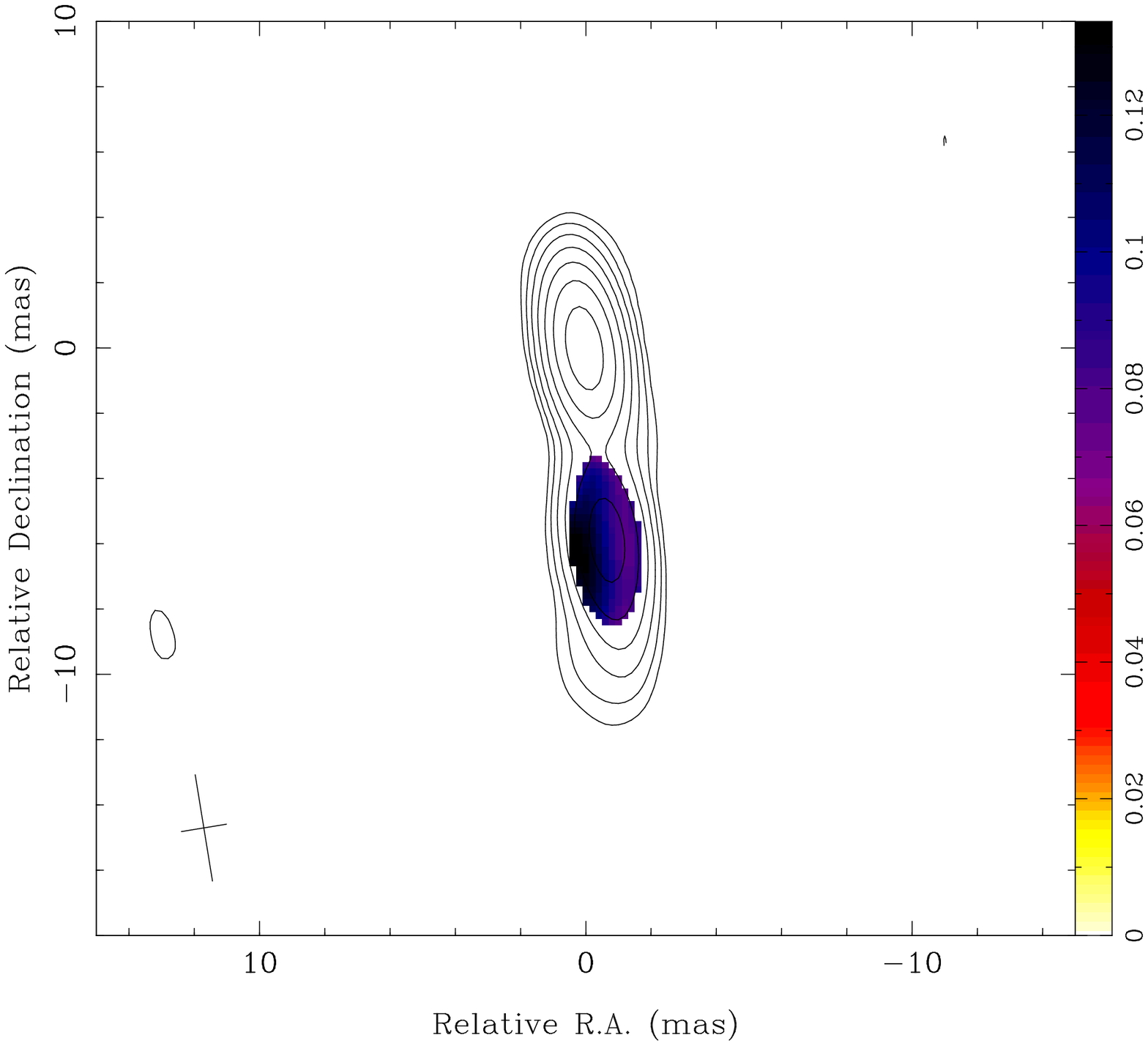}}}
      \quad

      \subfigure
      {\scalebox{0.4}{\includegraphics{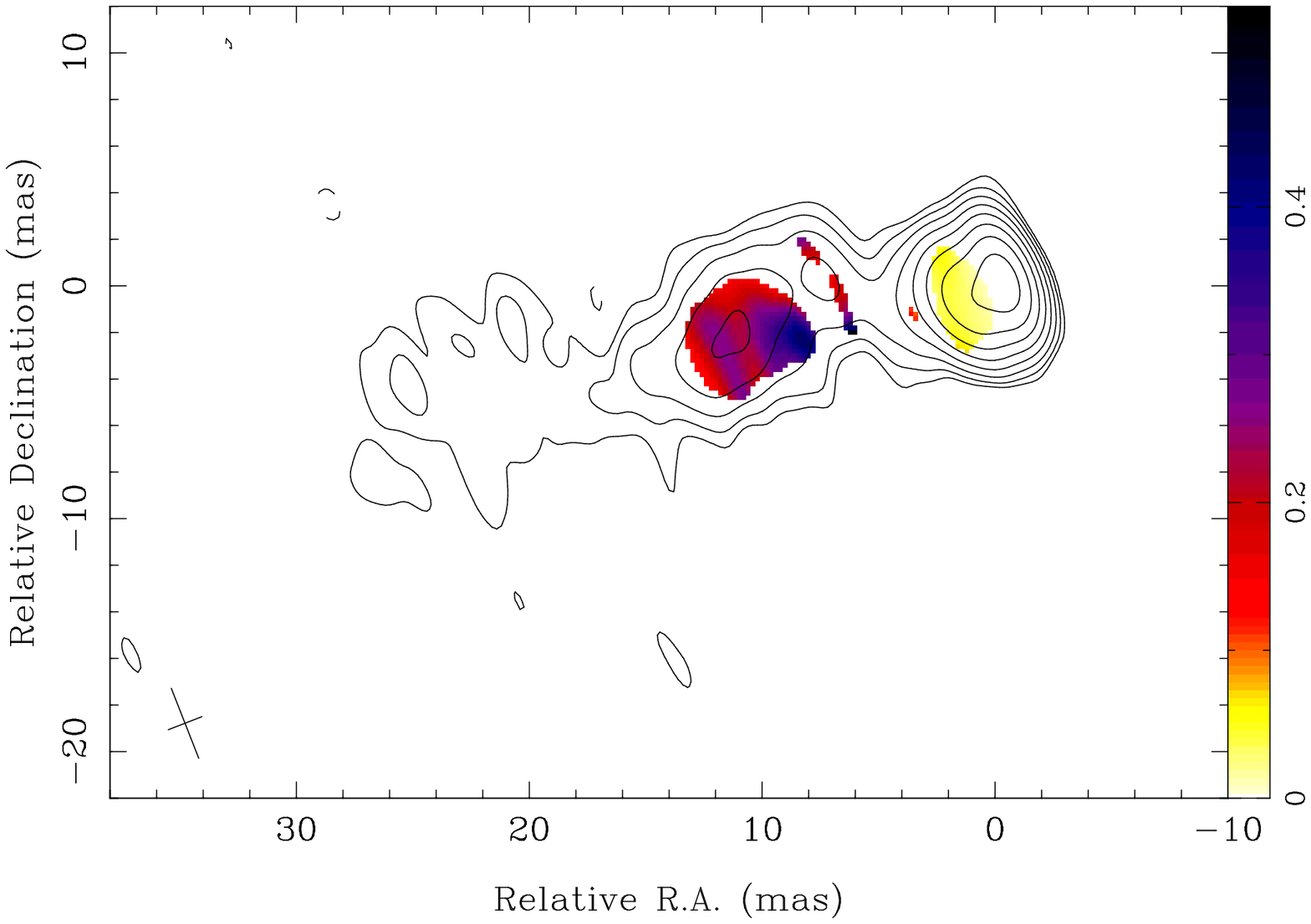}}}
      \quad}

    \mbox{
     \subfigure
     {\scalebox{0.34}{\includegraphics{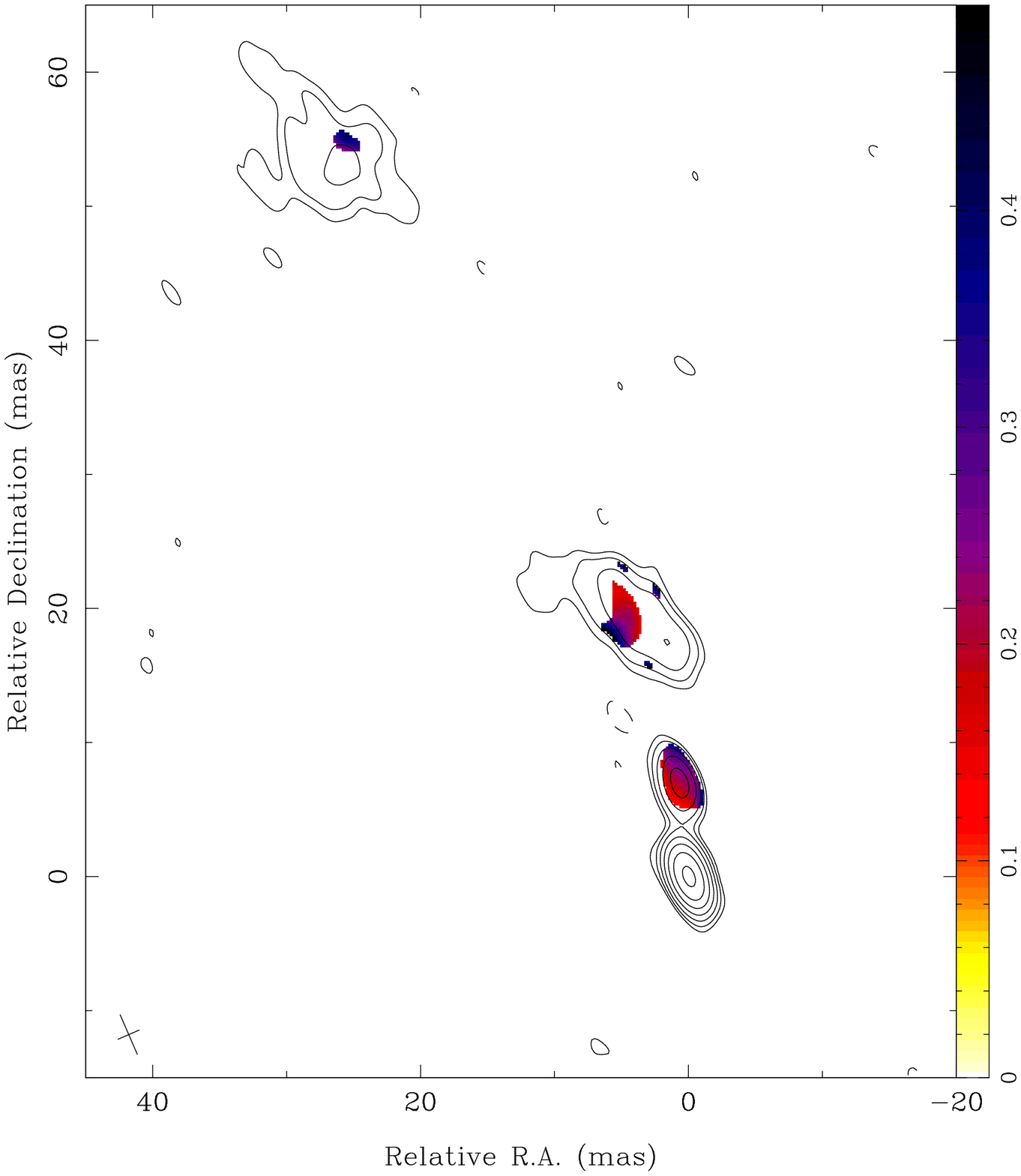}}}
      \quad}
  \end{flushleft}
  \caption{Fractional polarization. Upper left - SDSS J081432.11+560956.6; upper right - SDSS J090227.16+044309.6; middle left - SDSS J130522.75+511640.3; middle right - SDSS J144318.56+472556.7; lower left - SDSS J154817.92+351128.0.
 In each panel, the contour represents the overall VLBA radio structure in total intensity. }
  \label{fp}
\end{figure}

The fractional polarization images are given in Fig. \ref{fp} for five objects. We find that the fractional polarization in the jets ($\sim 10 - 30$ \%) is evidently higher than that of the radio cores ($< 5$ \%) (see Fig. \ref{fp}). This may plausibly be caused by the strong Faraday rotation
in the nuclear region due to the high plasma density, and/or depolarization due to the complex core structure. The polarization is clearly asymmetric and 
there are fractional polarization variations in some regions of the jet. This is likely caused by the relatively strong jet-ISM interaction at the locations of 
high fractional polarization due to the jet bulk motion to that side, in the cases of SDSS J081432.11+560956.6, SDSS J090227.16+044309.6, SDSS J130522.75+511640.3, and SDSS J154817.92+351128.0.
In SDSS J144318.56+472556.7, the polarization shows a prominent gradient along the jet direction, which well matches the jet knots. In this case, the surface shock may
play an important role in the emission of these regions. 

\subsection{Individual Objects}
\label{ind.obj.}

Besides the VLBI radio structure, the radio spectrum is also important to study the source nature. The individual objects were investigated in detail by collecting the multi-frequency and multi-epoch radio data from the NASA/IPAC Extragalactic Database (NED), in combination with the analysis on our VLBA images and archival VLBA data (if available). All sources were conventionally classified into flat-spectrum ($\alpha>-0.5$, in seven sources) and steep-spectrum ($\alpha\le-0.5$, in seven sources) RLNLS1s (see Table \ref{source}), by using the estimated radio spectral index $\alpha$ from low-resolution NED data (i.e. non-VLBI data), unless otherwise stated (see e.g., Sections \ref{0953} and \ref{1037}).

\subsubsection{SDSS J081432.11+560956.6}

\begin{figure}
  \begin{center}
    \mbox{
     \subfigure
     {\scalebox{0.4}{\includegraphics{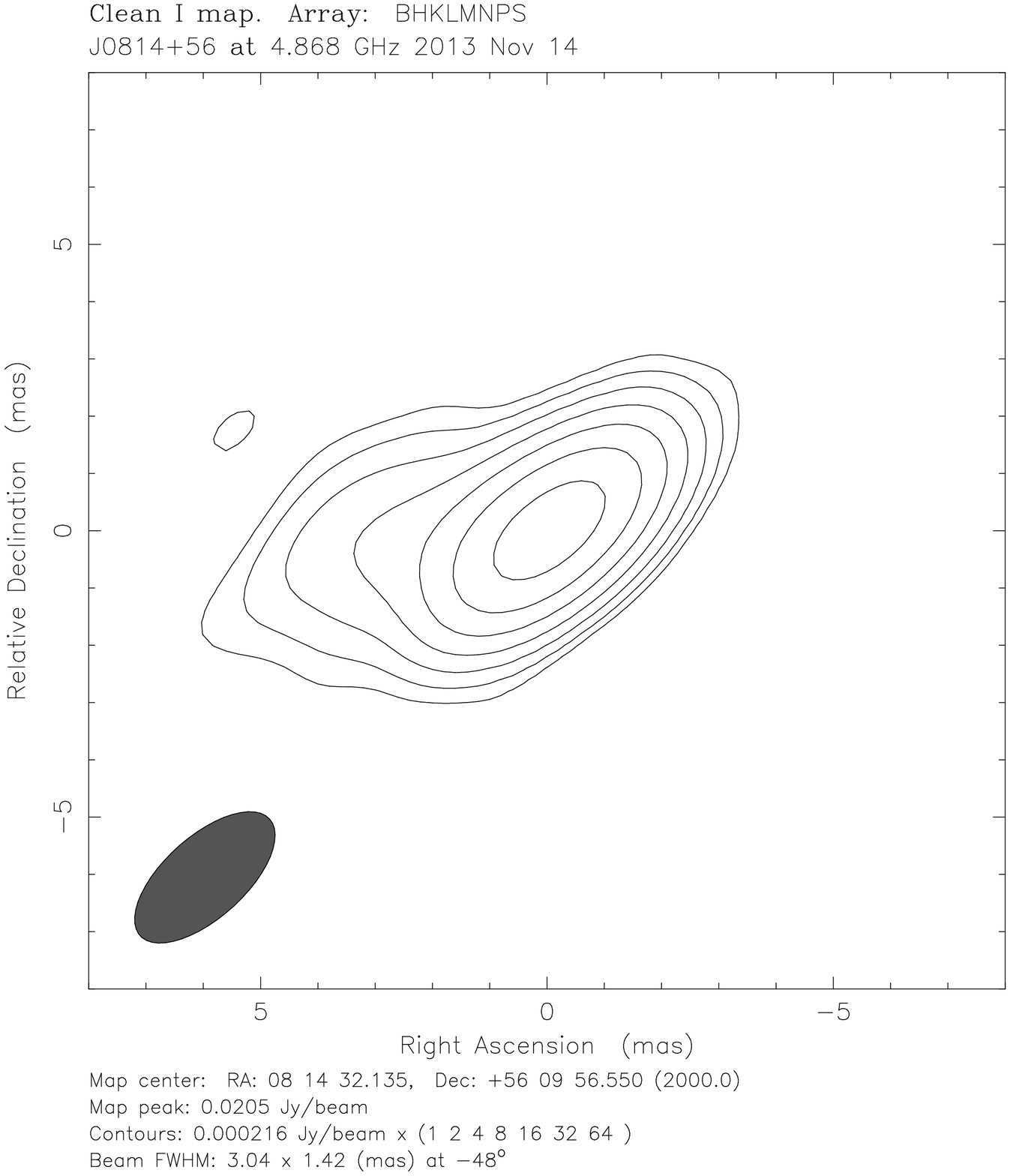}}}
      \quad


\subfigure
      {\scalebox{0.4}{\includegraphics{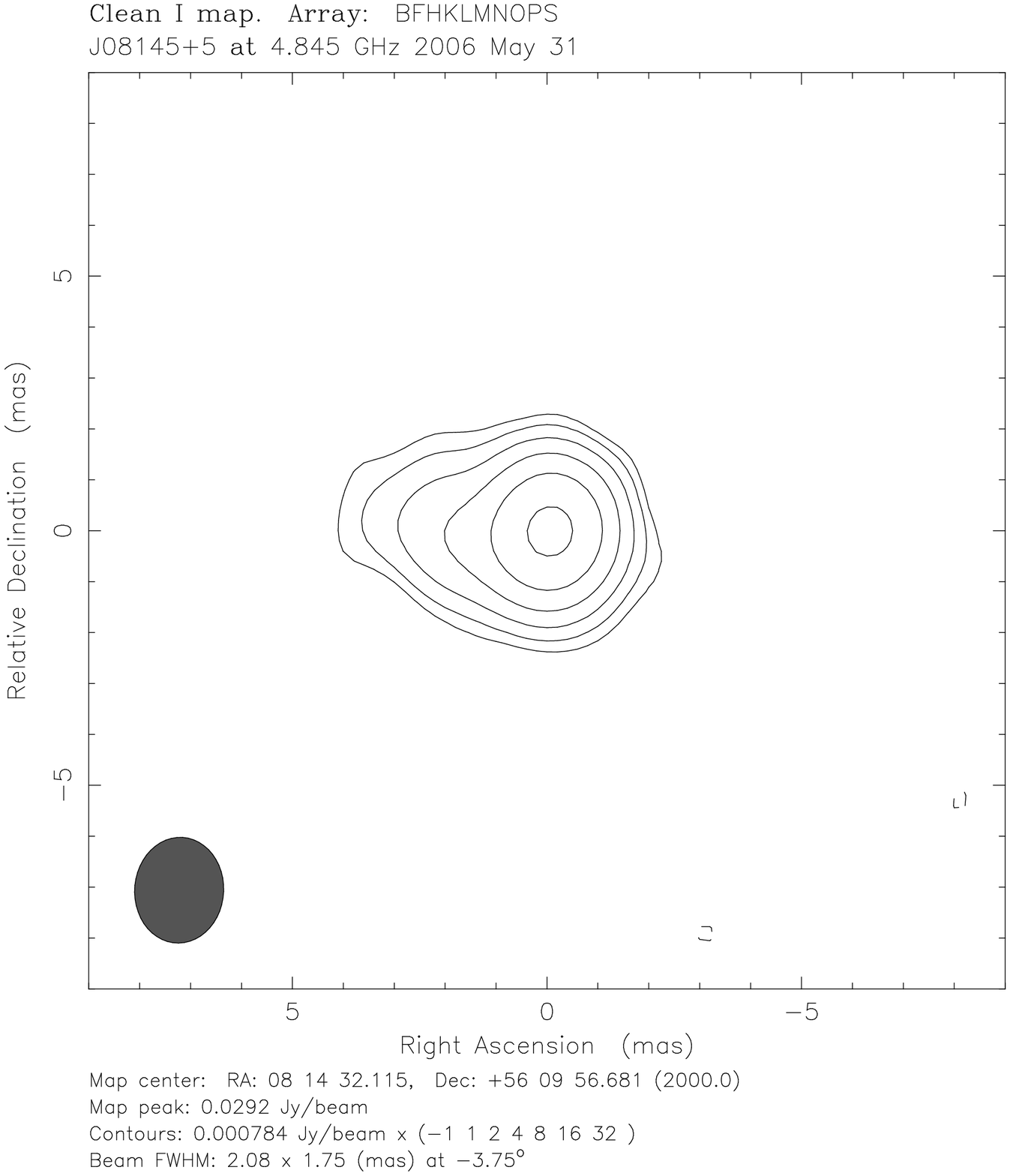}}}
      \quad}

    \mbox{
     \subfigure
     {\scalebox{0.4}{\includegraphics{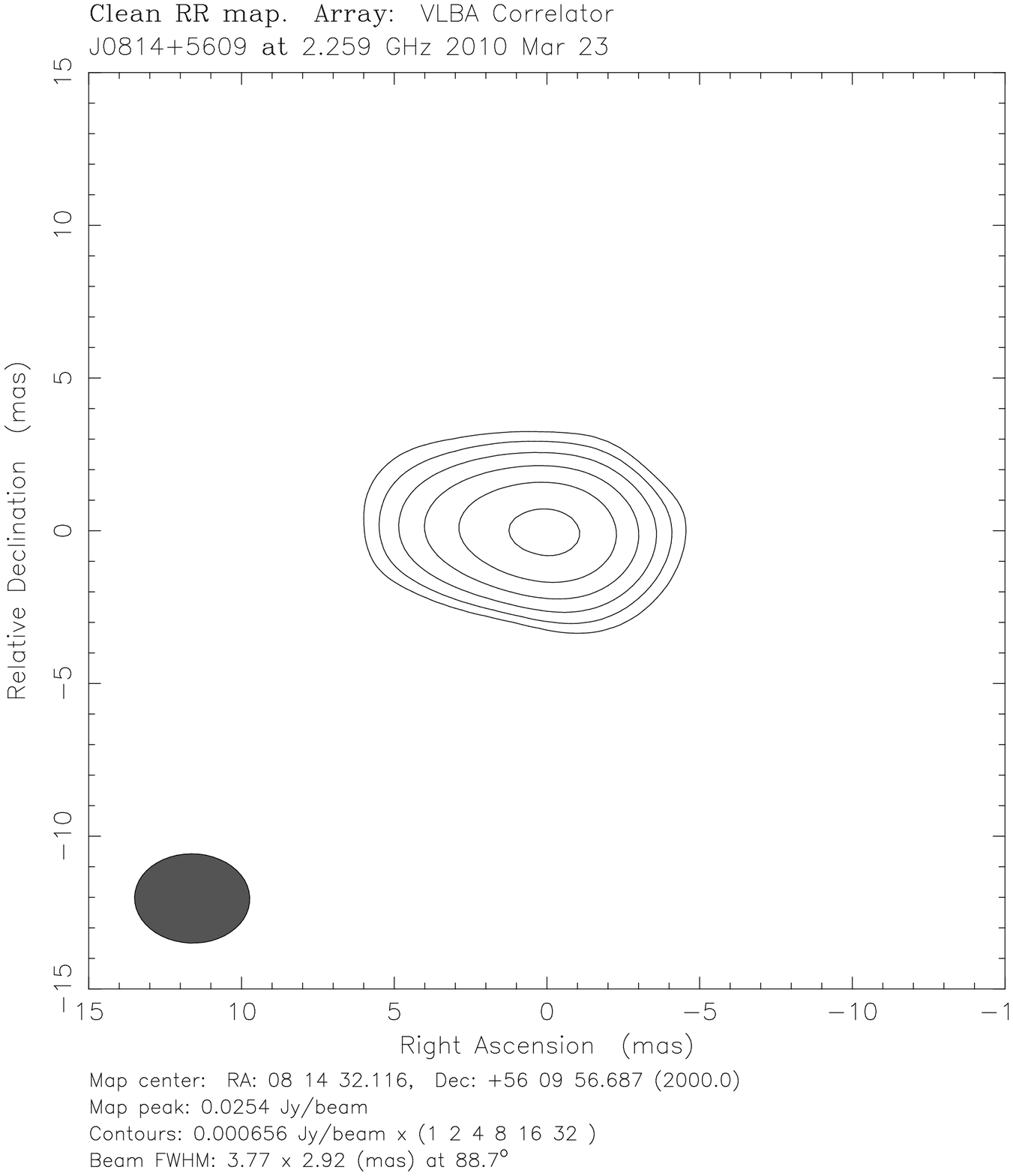}}}
      \quad

      \subfigure
      {\scalebox{0.4}{\includegraphics{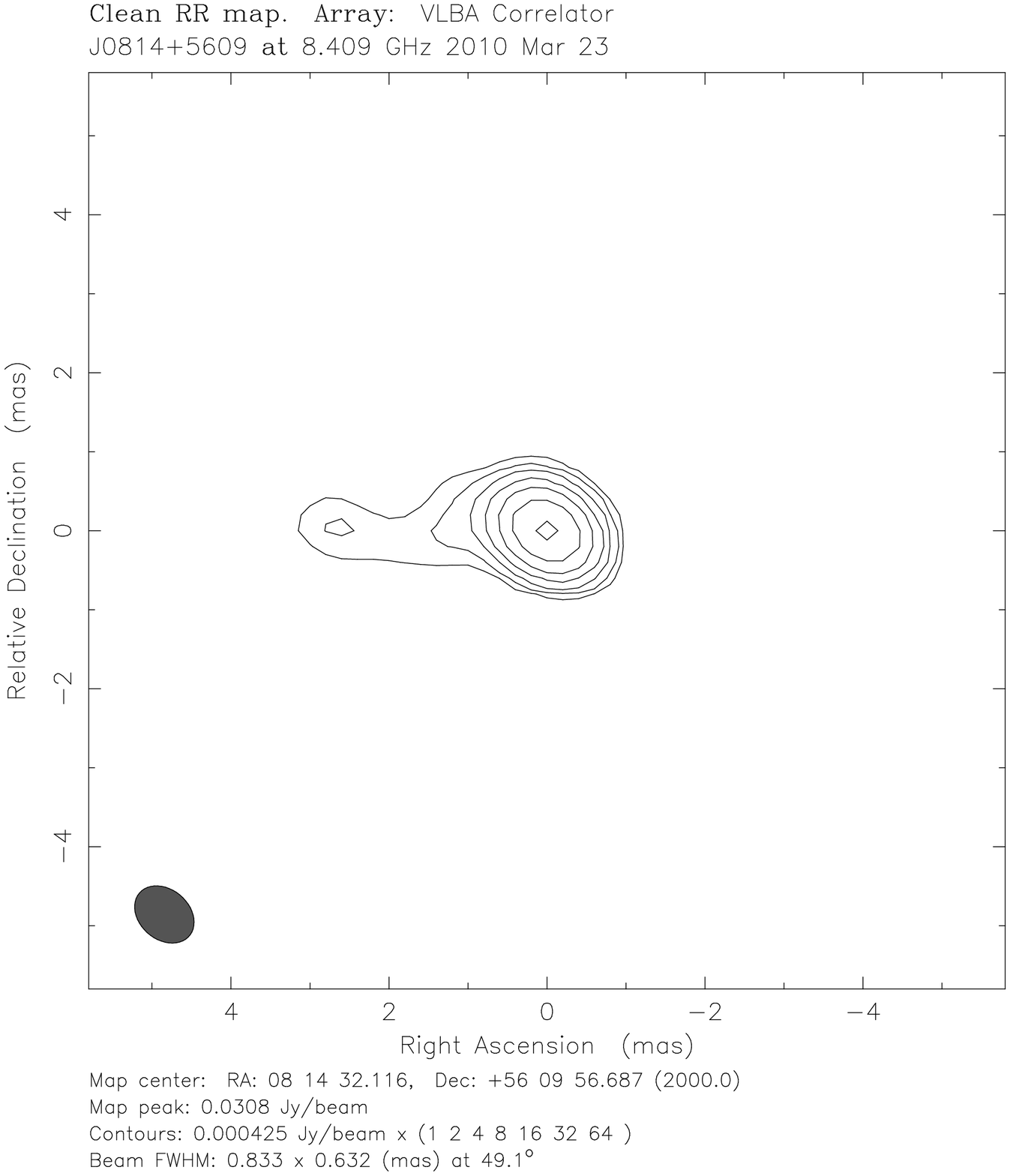}}}
      \quad}

  \end{center}
  \caption{VLBA 5 GHz image of SDSS J081432.11+560956.6, and archival 2.3, 5 and 8.4 GHz VLBA images.} \label{j0814}
\end{figure}

The source has a flat radio spectrum with a spectral index of $\alpha=-0.12$ based on 325 MHz, 1.4, and 4.85 GHz data collected from NED. It could therefore be a blazar-like object. Our 5 GHz image shows a core-jet structure with a weak jet component. The core is tentatively identified as the brightest component with an inverted spectrum of $\alpha=0.40$ between 5 and 6.7 GHz from our data. The simultaneous VLBA images at 2.3 and 8.4 GHz are presented in Fig. \ref{j0814} by using the available visibility data\footnote{http://astrogeo.org/obrs/obrs2\underline{ }cat.html} \citep{pet13}. The jet not visually apparent at 2.3 GHz is well resolved at 8.4 GHz. The core is confidently recognized as the brightest component with an inverted spectrum of $\alpha=0.18$ from the simultaneous 2.3 and 8.4 GHz data (see Table \ref{mfarchive}), confirming the core identification at 5 GHz. In Fig. \ref{j0814}, we also show the VLBA 5 GHz image from the archive, which was observed on May 31, 2006. A similar core-jet structure is seen. The core exhibits strong flux variability at 5 GHz when comparing with our observations, from which the variability brightness temperature can be calculated \cite[e.g.][]{yua08} to be $10^{10.3}$ K, indicating no strong beaming effect, consistent with the low brightness temperature $10^{10.1}$ K from our 5 GHz image (see Table \ref{modelfit}). However, we emphasize that the core brightness temperatures of $10^{11.8}$ and $10^{11.5}$ K at 2.3 and 8.4 GHz (see Table 3) are much higher, and therefore put much tighter constraints on the true brightness temperature. They do not yet exceed the inverse Compton catastrophe brightness temperature, though.

\subsubsection{SDSS J085001.17+462600.5}

The source was detected in the Giant Metrewave Radio Telescope LBDS-Lynx Region 150-MHz Radio Source Catalog \citep{ish10} and with the Westerbork Northern Sky Survey (WENSS) at 325 MHz \citep{ren97}, and it is compact in the FIRST image. The overall radio spectrum from 150 MHz to 1.4 GHz is steep with $\alpha=-0.62$. Therefore, the source is CSS-like. The spectrum is steep with a spectral index of $-0.73$ between 325 MHz and 1.4 GHz. However it changes to flat with $\alpha=-0.34$, between 150 and 325 MHz, implying a turnover. The object has only a compact component in our VLBA image (see Fig. \ref{j0850}), which can be tentatively treated as core with an inverted spectrum between 5 and 6.7 GHz. We measure a brightness temperature of $10^{9.8}$ K.

\subsubsection{SDSS J090227.16+044309.6}

The spectrum between 1.4 GHz and 5 GHz is flat with $\alpha=-0.24$ based on NED data. The source is resolved into core-jet structure, with the jet pointing towards the north-east at a position angle of about $60^{\circ}$ (see Fig. \ref{j0902}). The core brightness temperature from the VLBA image is $10^{11.4}$ K. The total VLBA flux density at 5 GHz ($\sim 102$ mJy) is very close to the GB6 flux (106 mJy), however only about half of that is from the radio core ($\sim50$ mJy).

\subsubsection{SDSS J095317.09+283601.5}
\label{0953}

The source is slightly resolved into a core-jet structure with a weak jet at position angle of about $33^{\circ}$ (see Fig. \ref{j0953}). The core brightness temperature is $10^{9.6}$ K. The source was only detected at 1.4 GHz, and not detected in the GB6 survey. Taking the detection limit of GB6, 18 mJy, as the upper limit, the spectral index between 1.4 and 5 GHz is expected to be steeper than $-0.7$, implying a CSS-like nature, which is tentatively supported by the steep spectrum between 1.4 GHz and VLBA 5 GHz (total flux) with $\alpha=-0.93$. However, it is hard to draw firm conclusion on the source nature with present data, and further observations are needed to study the object in detail.

\subsubsection{SDSS J103727.45+003635.6}
\label{1037}

The source was only detected at 1.4 GHz, so it is not possible to determine a radio spectral index with archival data. Combining the 1.4 GHz measurement with our new, non-simultaneous VLBA 5 GHz flux density, yields a spectral index of about $\alpha\sim-0.20$. This source then could be a flat spectrum object. The VLBA flux is about 22 mJy, therefore, it could have been detected in the GB6 survey. The non-detection according to the GB6 catalog, then indicates that the source underwent variability, consistent with the flat radio spectrum. Only a compact component is detected in our VLBA image (Fig. \ref{j1037}), with brightness temperature of $10^{10.5}$ K. 

\subsubsection{SDSS J104732.68+472532.1}
\label{1047}

\begin{figure}
  \begin{center}
    \mbox{
     \subfigure
     {\scalebox{0.4}{\includegraphics{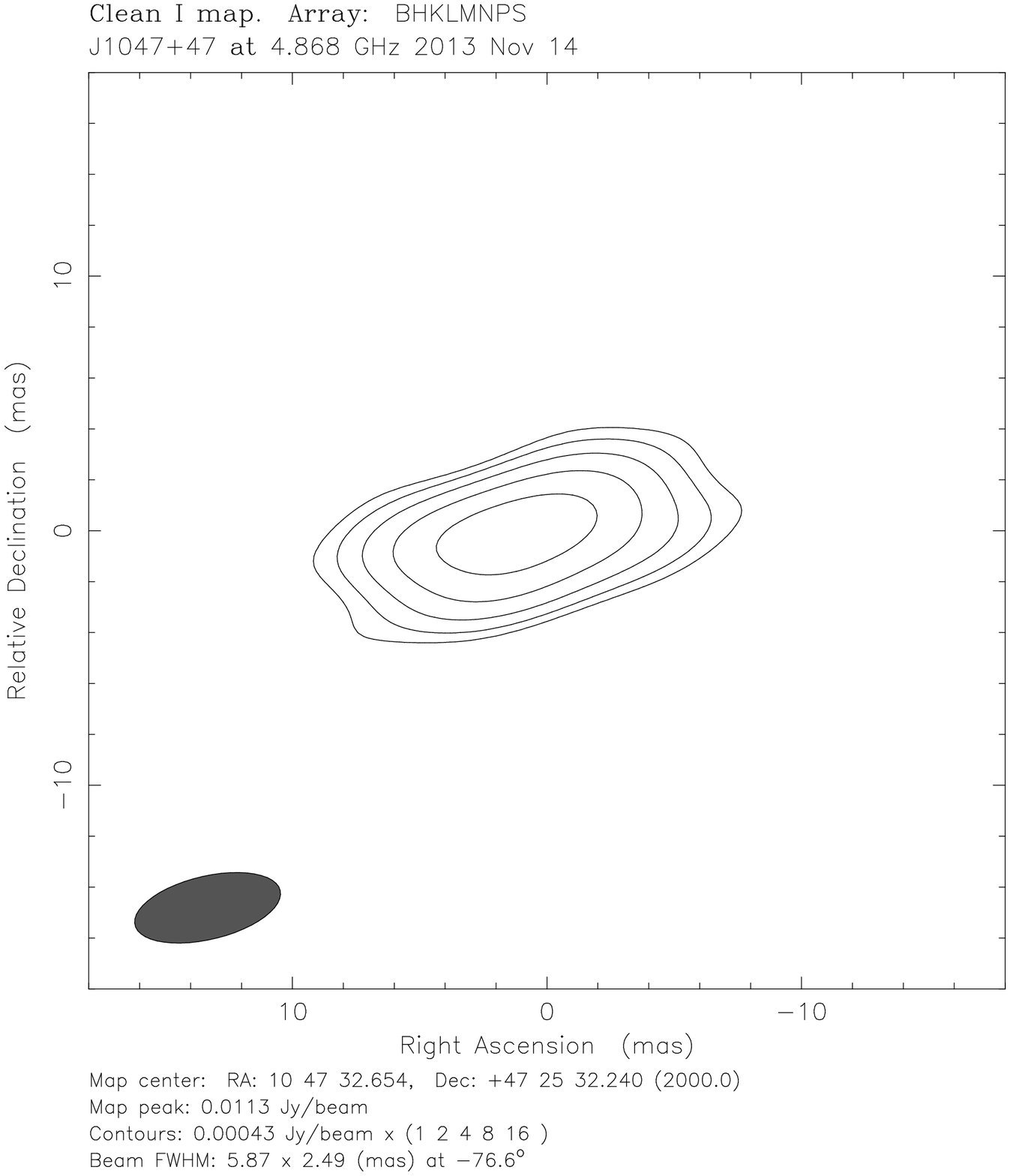}}}
      \quad

\subfigure
      {\scalebox{0.4}{\includegraphics{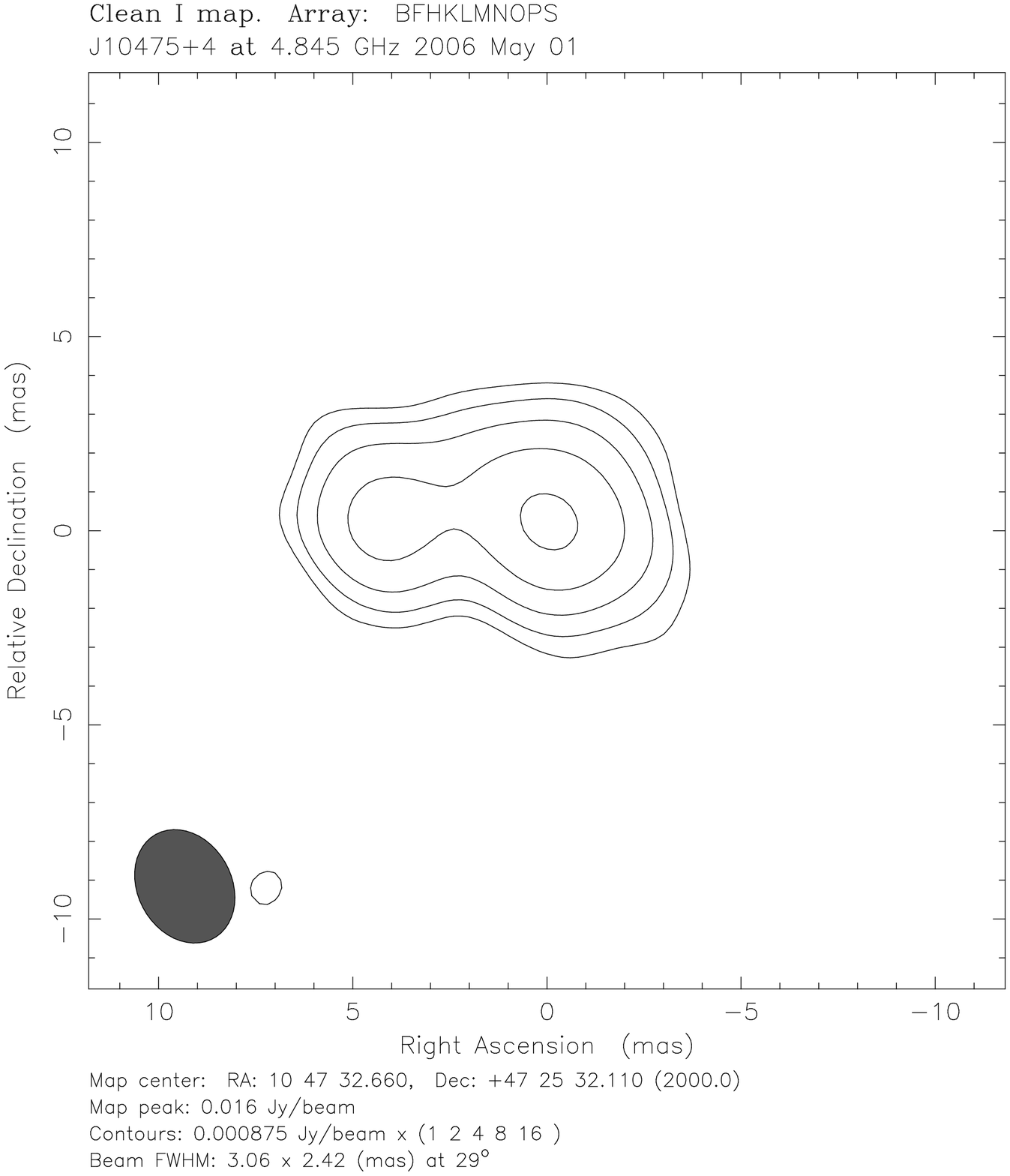}}}
      \quad}

    \mbox{
     \subfigure
     {\scalebox{0.4}{\includegraphics{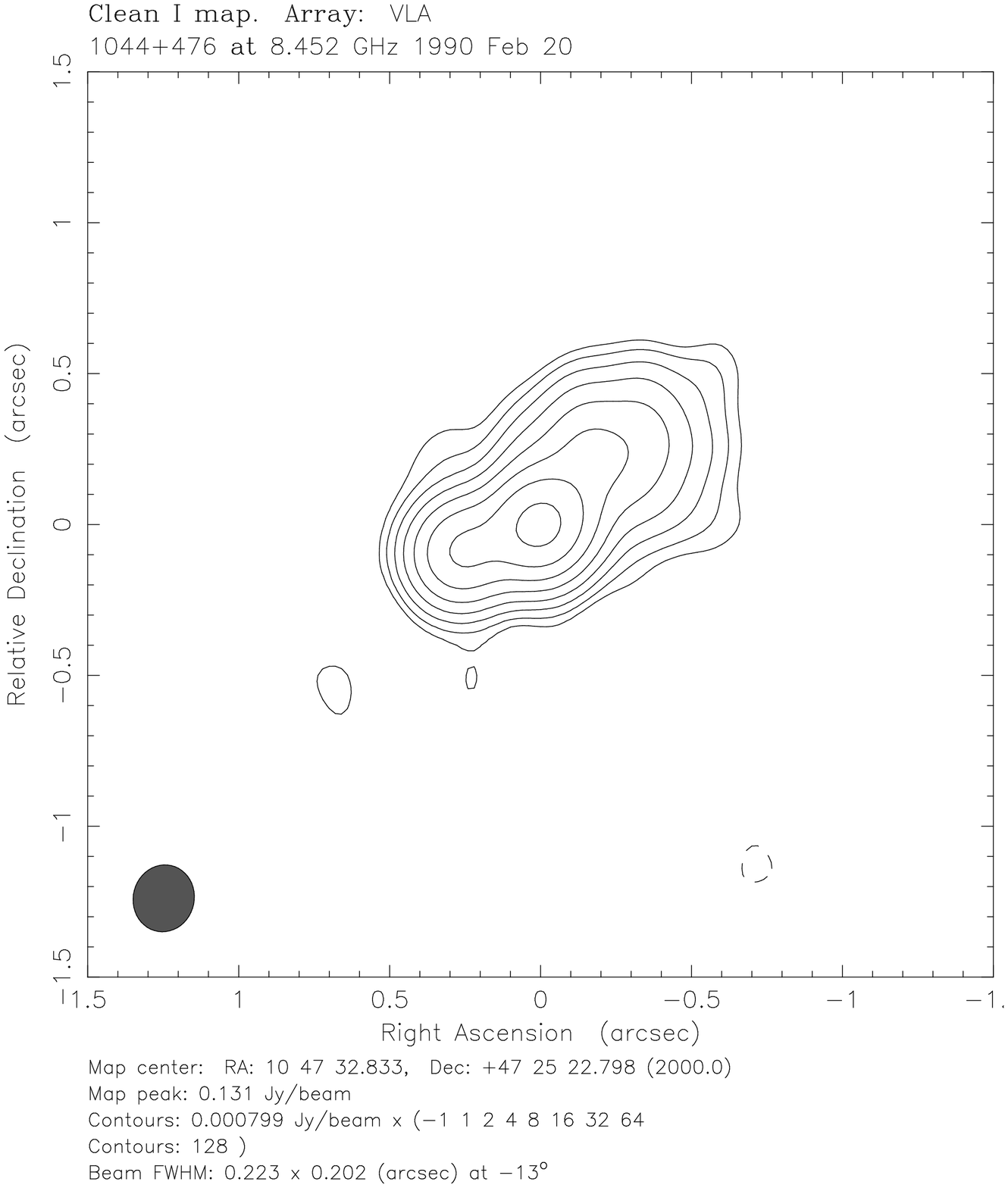}}}
      \quad

\subfigure
      {\scalebox{0.4}{\includegraphics{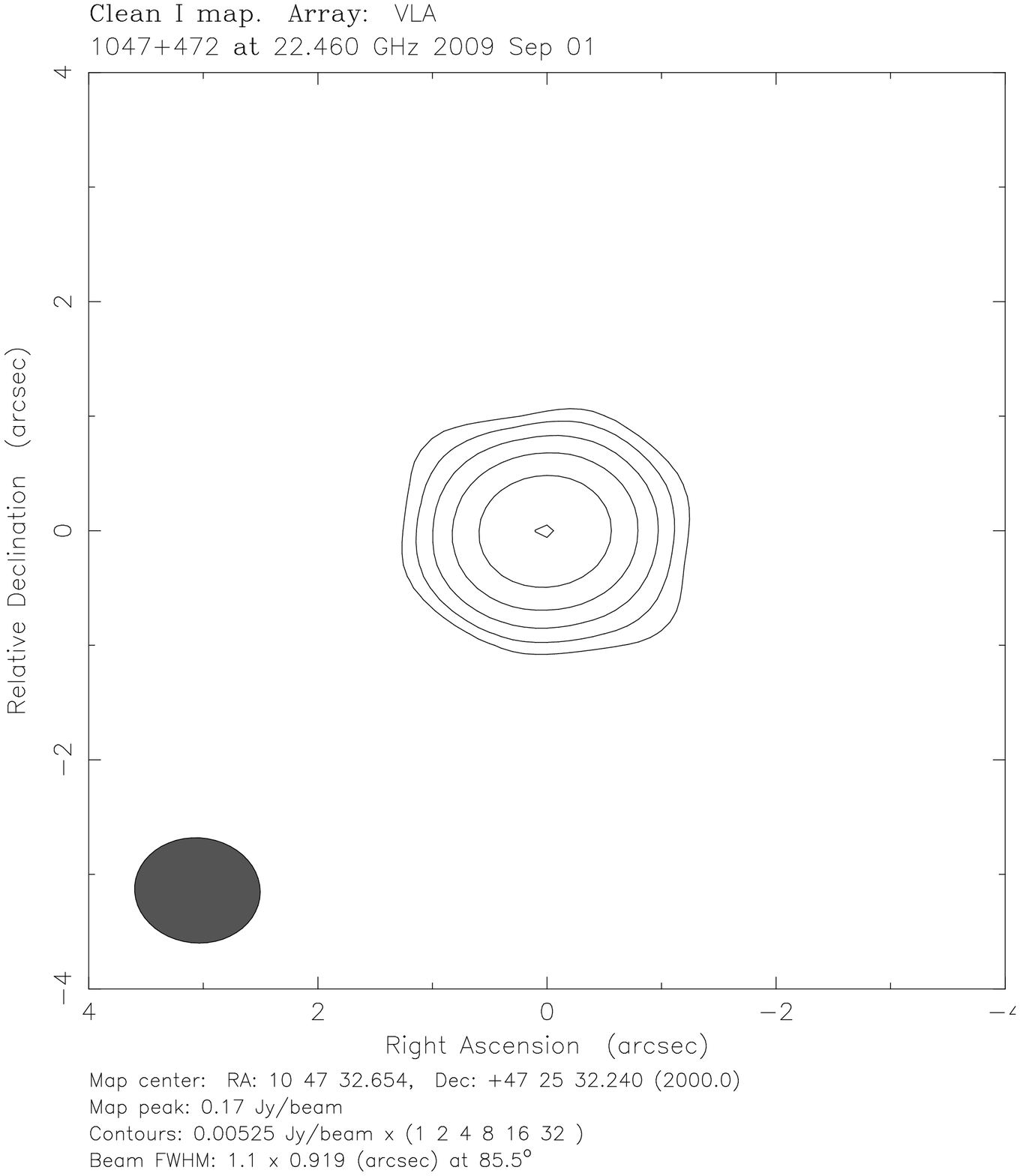}}}
      \quad}

  \end{center}
  \caption{VLBA 5 GHz image of SDSS J104732.68+472532.1, archival VLBA 5 GHz image, and VLA images at 8.4 and 22.4 GHz.} \label{j1047}
\end{figure}

The source is unresolved in the FIRST image, with upper limits on its extent of 1.16 and 1.14 arcsec for the major and minor axis, respectively, corresponding to a linear size of $\sim8.7$ kpc. The radio spectrum shows a clear trend for a turnover at low frequencies (see Fig. \ref{ri}), with a spectral index of $\alpha=-0.12$ between 74 and 151 MHz, while it is steep ($\alpha=-0.53$) at frequencies above 1.4 GHz. Such break frequencies at several hundred MHz in the source rest frame are often seen in small-scale radio sources, such as in CSSs \citep{ode98}. This object thus resembles CSS sources in terms of the compact size and steep spectrum. We found the object is unresolved in our VLBA image, however a core-jet structure is obtained from archival VLBA 5 GHz data because of higher resolution at east-west direction (see Fig. \ref{j1047}). Intriguingly, the GB6 5 GHz flux density is about a factor of 20 higher than the total flux density of our VLBA image. The flux difference can either be due to source variability, however this is unexpected in CSSs \citep{ode98}, or else, we miss a significant fraction of the extended flux. For further investigations, we analysed the VLA images at 8.4 and 22.4 GHz (see Fig. \ref{j1047}). A structure with two-sided lobes and a central core is clearly seen at a resolution of about 0.2 arcsec at 8.4 GHz, with a flux density ratio of about 1:1 between two lobes and the core (see Table \ref{mfarchive}). This implies that the source might be at large viewing angle, consistent with the steep spectrum. In contrast, only an unresolved compact component is present at 22.4 GHz likely due to the poor resolution, much lower sensitivity, and/or steep spectrum of the extended jet emission. But still, compared with the expected VLA core flux density at 5 GHz even assuming a spectral index of 0.0 between 5 and 8.4 GHz, our 5 GHz flux density are still much smaller. If the flux density difference is not due to strong variability, then a high fraction of the flux density needs to be resolved out, possibly distributed in extended lobes in sizes between sub-arcsec and mas. The core brightness temperatures are $10^{8.4}$ K, and $10^{9.0}$ K for our, and archival 5 GHz images, respectively, consistent with a large viewing angle.

\subsubsection{SDSS J111005.03+365336.3}

Only a compact component is detected in our VLBA image (see Fig. \ref{j1110}), with brightness temperature of $10^{10.0}$ K. The spectral index between WENSS 325 MHz and 1.4 GHz is steep, with $\alpha=-0.72$, making the source CSS-like. The spectrum remains steep, with $\alpha=-0.67$ between 1.4 and VLBA 5 GHz. 

\subsubsection{SDSS J113824.54+365327.1}

Only a compact component is detected in our VLBA image (see Fig. \ref{j1138}), with a brightness temperature of $10^{9.5}$ K. The spectral index between WENSS 325 MHz and 1.4 GHz is steep with $\alpha=-0.53$. 

\subsubsection{SDSS J124634.65+023809.0}

Only a compact component is detected in our VLBA image (Fig. \ref{j1246}), with brightness temperature of $10^{9.7}$ K. The spectral index between FIRST/NVSS 1.4 GHz and GB6 5 GHz is inverted ($\alpha=0.17$), making the source blazar-like. The VLBA flux ($\sim9$ mJy) is much less than that of the GB6 catalog (46 mJy). This flux difference can be either due to flux variations, or else most of flux is resolved out at VLBA resolution.

\subsubsection{SDSS J130522.75+511640.3}

\begin{figure}
  \begin{center}
    \mbox{
     \subfigure
     {\scalebox{0.4}{\includegraphics{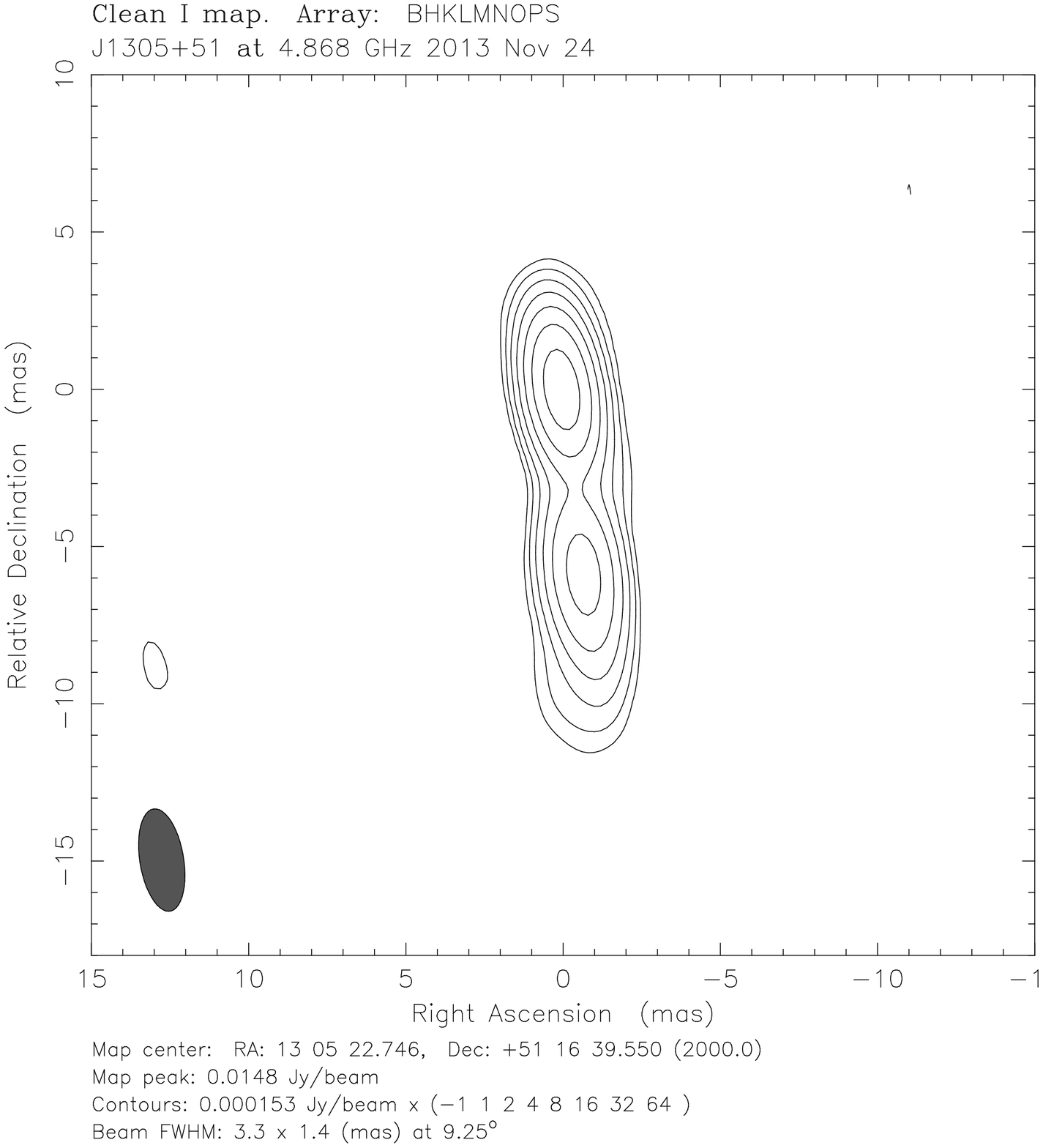}}}
      \quad

}

    \mbox{
     \subfigure
     {\scalebox{0.4}{\includegraphics{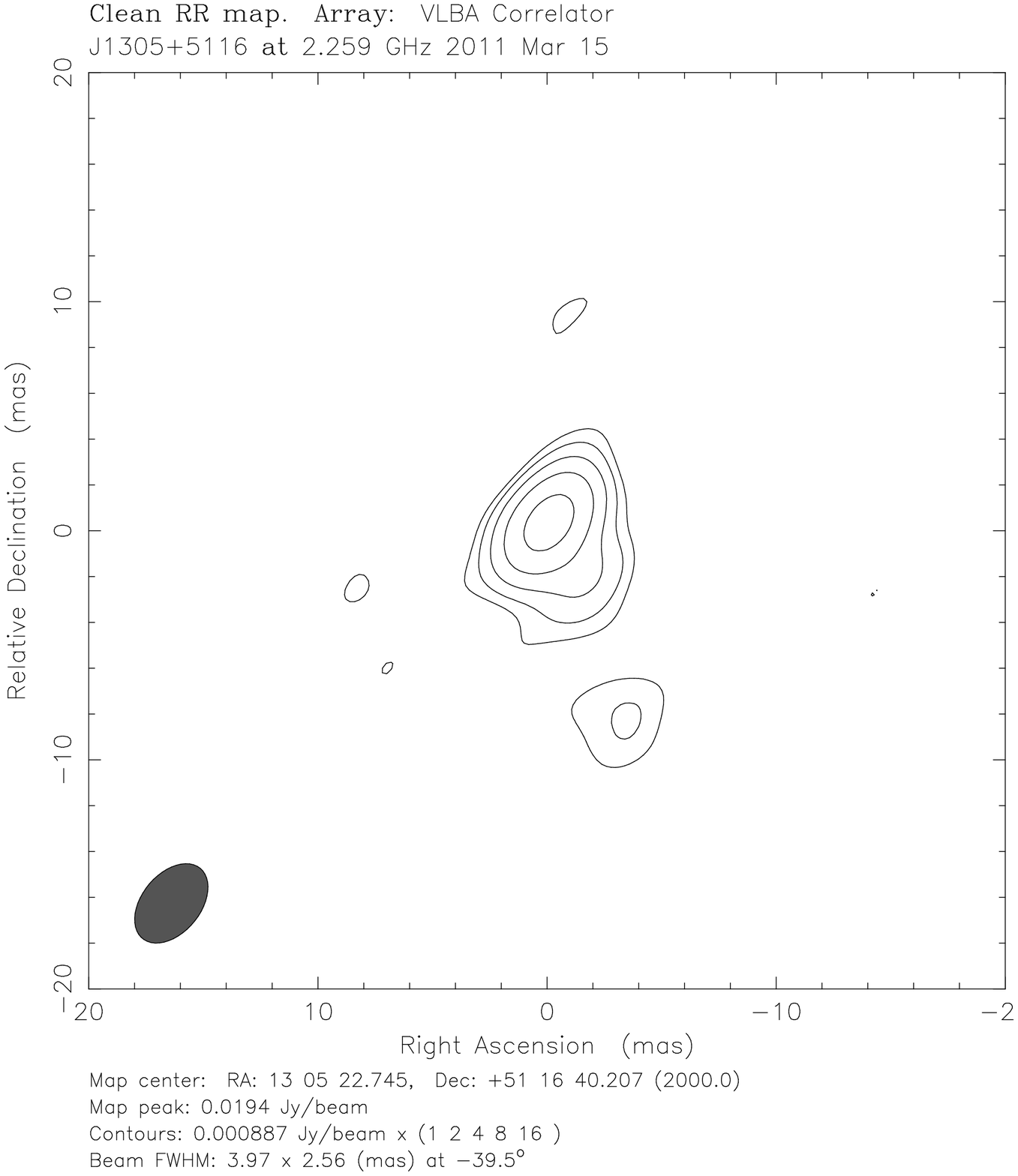}}}
      \quad

      \subfigure
      {\scalebox{0.4}{\includegraphics{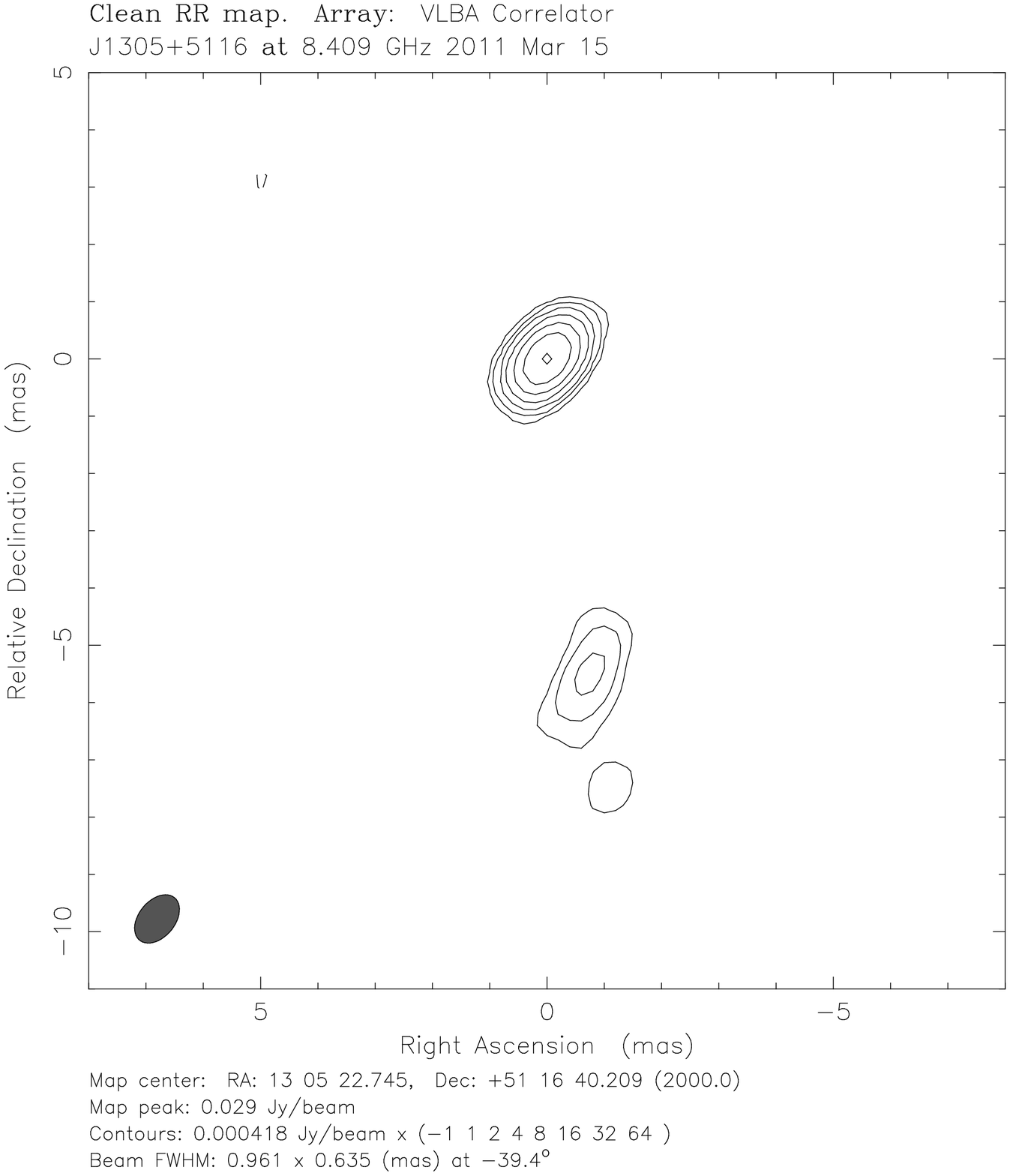}}}
      \quad}
  \end{center}
  \caption{VLBA 5 GHz image of SDSS J130522.75+511640.3, and archival VLBA 2.3 and 8.4 GHz images.} \label{j1305}
\end{figure}

The spectral index based on low resolution radio data is steep with $\alpha=-0.56$ from 151 MHz to 4.85 GHz (see Fig. \ref{ri}). In combination with the compact FIRST structure (1.34 arcsec, i.e. 10 kpc), this indicates that this source is CSS-like. The source is resolved into two components in our 5 GHz image (see Fig. \ref{j1305}). The northern component is brighter than the southern with a flux ratio of 1.7:1. The structure resembles that of compact symmetric objects, 
e.g. OQ 208 \citep{wu13}. However, the spectral index between 5 and 6.7 GHz is flat ($\alpha=-0.41$) for the northern component, while steep ($\alpha=-1.39$) for the southern. This implies
a core-jet structure for this source, rather than two steep-spectrum mini-lobes as in OQ 208. 
The core-jet structure is also supported by the inverted core spectrum of $\alpha\sim$ 0.29 between simultaneous VLBA 2.3 and 8.4 GHz, by analysing the available visibility data \citep{pet13}. The source is more resolved at 2.3 and 8.4 GHz due to the higher resolution in north-south direction (see Fig. \ref{j1305}). The total VLBA 5 GHz flux density ($\sim 23.9$ mJy) is only about half of that measured with GB6 (46 mJy), implying the other half is in the size between VLBA mas and GB6 resolution ($\sim3.5$ arcmin) if the flux difference is not due to variability. The core brightness temperature is about $5.8\times10^{10}$ K at 5 GHz. In contrast, the brightness temperature is about $10^{11.2}$ K at both 2.3 and 8.4 GHz. 

\subsubsection{SDSS J142114.05+282452.8}

The spectrum between FIRST/NVSS 1.4 GHz and GB6 5 GHz is flat with $\alpha= -0.20$. The source is slightly resolved into a core-jet structure with a weak jet component. 
The simultaneous VLBA images at 2.3 and 8.4 GHz are presented in Fig. \ref{j1421} by using the available visibility data \citep{pet13}. The weak jet is also seen in the 2.3 GHz image, however, it is not present in the 8.4 GHz image. The core is identified as the brightest 2.3 GHz component with a flat spectrum of $\alpha=-0.33$ between 2.3 and 8.4 GHz. The core brightness temperature is about $9.5\times10^{10}$ K at 5 GHz, and $10^{10.6}$ and $10^{10.1}$ K at 2.3 and 8.4 GHz, respectively.

\subsubsection{SDSS J144318.56+472556.7}

\begin{figure}
\epsscale{1.}
\plotone{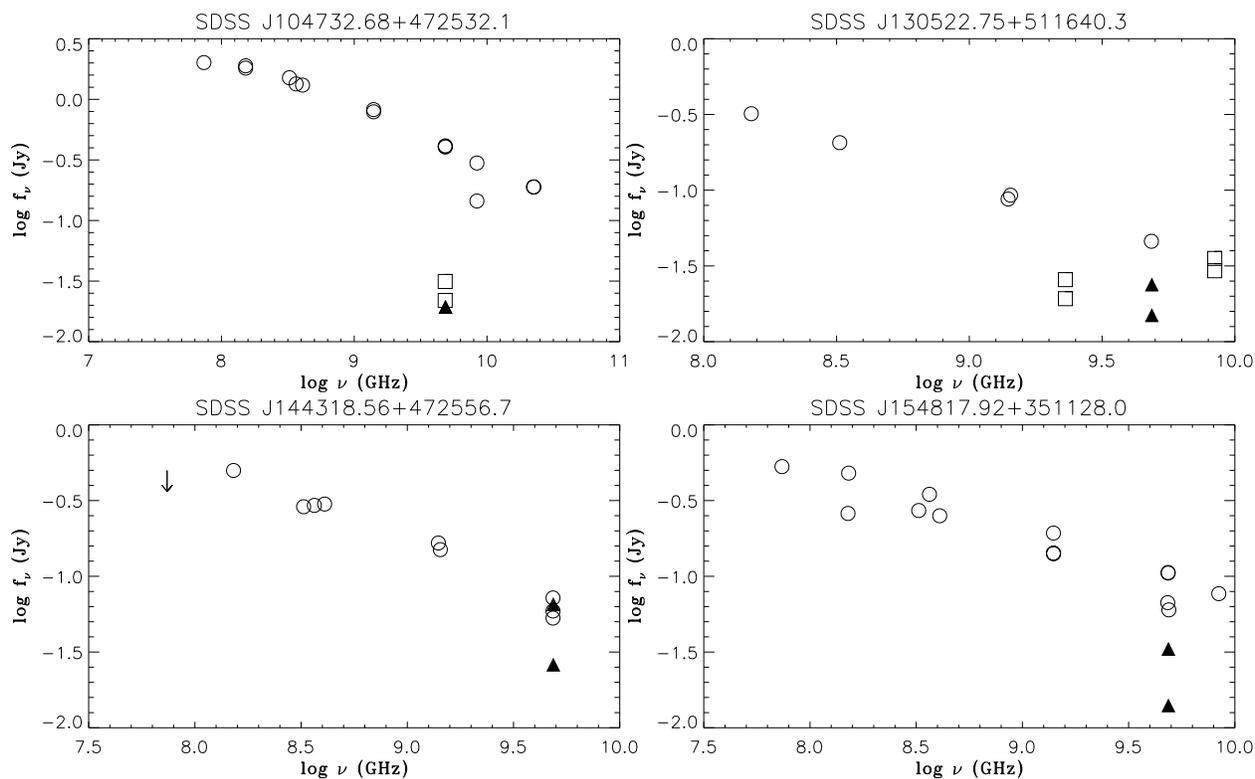}
\caption{Radio spectra. The circles are based on NED data. The solid triangles are core and total VLBA flux densities from our measurements at 5 GHz. The rectangles are core and total
VLBA flux densities from measurements on the archival data. The downward arrow is the upper limit of the flux density at 74 MHz. \label{ri}}
\end{figure}

The multi-frequency NED data show that the spectral index between 151 MHz and 4.85 GHz is steep with $\alpha=-0.60$ (see Fig. \ref{ri}). The source is not listed in the Very Large Array Low-frequency Sky Survey Redux catalog \cite[VLSSr,][]{lan14}, which only contains the sources with 74 MHz flux density at greater than $5\sigma$ significance, i.e. $\rm \sim 0.5~ Jy~ beam^{-1}$. Taking 0.5 Jy as the upper limit at 74 MHz, the radio spectrum shows a clear trend of a turnover toward low frequencies (Fig. \ref{ri}). Indeed, the spectral index is $-0.57$ between 151 and 408 MHz, while it is $-0.77$ between 1.4 and 4.85 GHz. This object is unresolved in the FIRST image with a size of 1.18 arcsec (major axis), corresponding to a linear size of 8.4 kpc. Given its steep spectrum and compact structure, this NLS1 shares similarity with CSS sources. The VLBA 5 GHz image is shown in Fig. \ref{j1443}. The source is resolved into seven components along the south-west direction within 15 mas ($\sim107$ pc) (see Table \ref{modelfit}), and diffuse emission is clearly detected at large distance up to 30 mas. Tentatively calculating the spectral index between 5 and 6.7 GHz from our VLBA data, we found that the spectra of all seven components are steep. This implies that the radio core is not directly observed, perhaps hidden in the brightest component. However, multi-frequency VLBI observations are needed before we can draw firm conclusions. The brightness temperature at 5 GHz is $10^{10.3}$ K for the brightest component. The total flux density at 5 GHz from our VLBA image is close to that in the single dish GB6 catalog (see Fig. \ref{ri}), which indicates that the source is very compact, and there is likely no much extended emission even at sub-arcsec scale. The extended and/or diffuse emission can be only resolved at mas scale, which contains a large fraction of the jet emission as the flux density of the brightest component is only about $\sim 40\%$ of the total flux density.

\subsubsection{SDSS J154817.92+351128.0}

The source is well resolved into a core-jet structure (see Fig. \ref{j1548}). The jet is along the south-north direction in the central region, then it turns to the north-west. At a large distance of about 60 mas ($\sim353$ pc), diffuse emission is evident. The core is identified with the brightest component, with an inverted spectral index of $\alpha=0.4$, tentatively estimated from our 5 and 6.7 GHz measurements. However, its 5 GHz flux density is only about 42\% of the total flux density. The source was detected at 74 MHz in the VLSSr catalog. The multi-wavelength data shows an overall flat spectrum ($\alpha=-0.43$) from 74 MHz all the way to 8.4 GHz, and the spectrum is flat ($\alpha=-0.45$) above 1.4 GHz (see Fig. \ref{ri}). The overall radio spectrum shows that this source is most likely a flat-spectrum source, resembling blazars. However, there perhaps is a turnover below 100 MHz, possibly due to synchrotron self-absorption. The total VLBA flux density (33 mJy) is about 55\% of that of VLA (60 mJy, from NED), indicating that about half of the jet emission is resolved out at VLBA resolution.
However, alternatively, the flux difference between VLBA and VLA can also be due to the flux variations since it is a flat-spectrum object. The brightness temperature of the core is $10^{9.9}$ K at 5 GHz. A variability brightness temperature of $10^{13}$ K was found at 4.85 GHz by \cite{yua08}, which implies a Doppler factor of $\delta_{\rm min}=2.2$.  

\subsubsection{SDSS J163401.94+480940.2}

Only a compact component is detected in our VLBA image (see Fig. \ref{j1634}), with brightness temperature of $10^{10.1}$ K. The spectrum between WENSS 325 MHz and NVSS 1.4 GHz data of similar resolution is flat with $\alpha=-0.47$. The FIRST flux density is only about 55\% of NVSS, implying that the remaining flux is either resolved out at FIRST resolution, or else the source is variable. 


\section{Discussion}


All our sources were selected with 1.4 GHz flux densities $\ge$10 mJy, and they are very radio loud with $R>100$ (see Table \ref{source}). Therefore, they represent the most extreme RLNLS1s in terms of their radio loudness $R$. However, their jet properties are not homogeneous, in terms of the mas scale morphology and the overall radio spectral shape. A variety of mas scale morphologies has been found, including compact core only (7 out of 14), compact core-jet (5 of 14), and extended core-jet structures (2 of 14). Based on the compactness defined as the core to total flux ratio, most sources (12 of 14) are compact, and core-dominated with compactness larger than 0.5. Two sources,  SDSSJ144318.56+472556.7 and SDSSJ154817.92+351128.0, have a compactness of $\sim0.4$ with a large-scale jet extending to about 30 and 60 mas (see Figs. \ref{j1443} and \ref{j1548}), respectively. A one-sided core-jet structure is detected in all the resolved objects (7 out of 14), which is commonly found in blazars. 
As for the radio spectral shape, we found flat spectra in seven sources, and steep spectra in the other seven sources, albeit large uncertainties in SDSS J095317.09+283601.5 and SDSS J103727.45+003635.6 (see Section \ref{ind.obj.} and Table \ref{source}). In our sample, there is no strong relationship between the compactness and spectral index. The steep spectrum sources do not prefer to have more extended structures, while those with flat spectra do not appear more compact. 

It should be noted that our measurements of brightness temperature are typically lower limits, and the true brightness temperatures can therefore be higher. However, taken at face value, the median brightness temperature of our sample is significantly lower than that of typical blazars \cite[of order $10^{12}$ K;][]{kov05,kov09}. If confirmed by future higher-resolution observations, this finding can, in principle, be explained in two ways. Either, our sources are not significantly beamed, then likely implying a low bulk speed of the jet. This is likely the case for flat-spectrum sources, since the viewing angles are expected to be small in these sources. Alternatively, these sources are seen at relatively large angle, and Doppler-deboosting plays a role, which could be the situation in steep-spectrum objects (e.g., SDSS J104732.68+472532.1, see Section \ref{1047}). Indeed, although not conclusive due to the lower limit, the brightness temperatures of steep spectrum sources are systematically lower than those of flat spectrum sources with median values of $10^{9.8}$ and $10^{10.1}$ K for steep and flat sources (see Fig. \ref{ccj}), respectively. This is qualitatively consistent with the expectation that flat spectrum sources are usually viewed at relatively smaller angles, and then have a relatively larger beaming effect than steep spectrum sources. Moreover, the steep-spectrum RLNLS1s were recently proposed as the parent population of flat-spectrum RLNLS1s \citep{ber15a}. In this scenario, the fundamental jet physics could be similar, such as the jet bulk speed, and the observational differences in two populations could be mainly due to the different viewing angle. Therefore, given the overall properties of our sources, the first possibility is favored that the jet bulk speed is likely low in our sources. Consistently, the radio monitoring of $\gamma$-ray-detected NLS1s shown that the computed variability brightness temperatures are in general moderate, in contrast to the majority of blazars, implying mildly relativistic jets \citep{ang15}. The mildly relativistic jets were also found in \cite{ric15} for three RLNLS1s based on various information, such as the moderate core brightness temperature, the jet/counter-jet intensity ratio and the extension length ratios from VLBA and VLA observations. 

Flux variability of each source was searched for by comparing the FIRST and NVSS fluxes. Due to the different beam sizes of NVSS and FIRST, we conservatively treat the sources with the FIRST peak flux larger than the integrated flux of NVSS as possible variable sources. The significance of the variation $\sigma_{\rm var}$ is calculated as the ratio of the flux difference between the FIRST peak flux and NVSS integrated flux to the combined uncertainties including the corresponding uncertainties in FIRST and NVSS fluxes and the additional systematic uncertainties \cite[see e.g.][]{wan06}. Using a threshold $\sigma_{\rm var}>3$, we found flux variations in only one source, SDSSJ085001.17+462600.5 ($\sigma_{\rm var}$=4.0). For most sources (10/14), the FIRST peak and integrated fluxes are in agreement with the integrated NVSS flux within 10\%. The mean values of the flux ratio of FIRST to NVSS are $0.99\pm0.17$ and $1.03\pm0.19$, for FIRST peak and integrated fluxes, respectively. This again shows that variations are usually not significant for most sources. Moreover, it indicates that all our sources are mostly compact, without much extended diffuse emission, as often found in RLNLS1s \citep{doi12,yua08}. Indeed, all fourteen NLS1 galaxies of our sample were detected as single unresolved sources within 1 arcmin of the SDSS position in both FIRST and NVSS, indicating compact radio structure at FIRST resolution (5 arcsec), except if any extended emission was faint, and beyond the detection threshold of NVSS and FIRST. We tentatively use the deconvolved major axis provided in the FIRST catalog as an upper limit of the linear size of each source. The linear sizes range from 2.1 to 17.4 kpc (see Table \ref{source}), confirming that they are compact.   

The association of RLNLS1s with CSS sources has already been proposed by many authors \citep{mor00,osh01,gal06,kom06}, and recently by \cite{cac14}. In addition, the hypothesis of CSS sources as the parent population of RLNLS1s was also discussed, in which the RLNLS1s with a flat radio spectrum could be CSS sources with jets pointing towards the observer \citep{yua08,cac14}. Indeed, the steep-spectrum RLNLS1s were recently found to likely be the best candidates for the parent population of flat-spectrum RLNLS1s, although not conclusive and disk-hosted radio galaxies have yet to be included \citep{ber15a}. In our sample, seven sources have compact structure and steep spectrum (although uncertain in the case of SDSS J095317.09+283601.5, see Section \ref{ind.obj.}). These share some resemblance with CSS sources. However, their VLBI structure is different from that of typical CSS sources \citep{ode98,dal13}. As shown in \cite{dal13}, the majority of CSS sources have a two-sided structure and the radio emission is dominated by lobes, jets and hotspots. In our sample, four out of seven steep spectrum sources have core-only morphology, and a one sided core-jet structure is found in the remaining three objects. For core-jet sources, the core flux dominates in two objects ($\sim90$\% in SDSSJ095317.09+283601.5, $\sim63$\% in SDSSJ130522.75+511640.3). Only in SDSSJ144318.56+472556.7, the radio emission is dominated by the elongated jet structure and diffuse emission with about 40\% of the total emission in the core. While most of the high-power typical CSS sources have a double or triple morphology at VLBI scales \citep{ode98,dal13}, compact core only and core-jet structures have been detected in CSS sources with relatively low radio power \citep{kun06,kun07}. In terms of the linear size (using FIRST major axis as upper limit) and 5 GHz radio power \citep{yua08}, our NLS1 galaxies resemble the low-power CSS sources of \cite{kun10}. In the radio power versus linear size diagram, the low-power CSS objects occupy the space below the main evolutionary path of radio objects. The authors argued that many of these might be short-lived objects, and their radio emission may be disrupted several times before they become FR IIs. Alternatively, some of the short-lived low-power CSS objects could be precursors to large-scale FR Is \cite[see also][]{an12}. The argument linking RLNLS1s with CSS sources mainly relies on the compact arcsec scale structure in RLNLS1s. The majority of RLNLS1s exhibit compact radio emission, while extended emission has only been detected in nine RLNLS1s \citep{doi12,ric15}. \cite{doi12} argued that the jet of RLNLS1s has low kinetic power because of the small mass and because it has to propagate in a gas-rich environment, which can be likely responsible for rare detection of extended emission. However, recent VLA observations of three arbitrarily selected RLNLS1s, show that all sources possess kpc-scale extended emission \citep{ric15}. One of the CSS-like sources in our sample, SDSS J095317.09+283601.5, was reported to have a two-sided,
mildly core-dominated jet structure with diffuse lobes at an overall projected linear size of about 70.2 kpc in their study, largely exceeding the typical size of CSS sources. The authors argued that the extended emission could be more common when observed with enhanced sensitivity, such as the upgraded VLA.  
While this indicates the capability of jet extension to kpc scale, more observations are needed to study the fraction of extended emission in the overall RLNLS1 population.  

With NVSS, polarized flux was detected in 11 out of the 14 NLS1 galaxies of our sample. The fractional polarization ranges from 0.5\% to 5.2\% with a mean value of 2.2\%. 
There is no significant systematic difference of fractional polarization between steep and flat spectrum sources. In the steep spectrum sources, the fractional polarization is similar to that of CSS sources \cite[$\sim1\% - 3\%$ at 6 cm, see][]{ode98}. Several studies of the polarization properties of CSS sources have shown that sources smaller than a few kpc have very low values of polarized emission at frequencies below 8.4 GHz \cite[e.g.][]{cot03}. This result has been interpreted by assuming that compact sources are highly depolarized by the dense interstellar medium that enshrouds the radio emission and acts as a Faraday screen. When using the deconvolved major axis as the upper limit of the source linear size, we found that the fractional polarization of steep spectrum sources is below 2\% for objects smaller than 10 kpc, while higher than 4\% for larger sources. This suggests that there may be an abrupt change in the properties of the ISM at a radius of approximately 10 kpc, possibly related to the outer boundary of the narrow line region (NLR), which however needs further studies. The radio power distribution of our RLNLS1 galaxies overlaps with the lower part of that of CSSs \citep{ode98}. As such, if there is an overlap between CSSs and RLNLS1 galaxies, it should consist mostly of CSSs having low radio power and relatively small black hole masses.

Besides seven steep spectrum sources in our sample, seven galaxies show flat radio spectra, implying relationships with blazars. Recently, \cite{fos15} conducted a multi-wavelength survey of 42 RLNLS1s with flat radio spectra. They found that the jet power is generally lower than FSRQs and BL Lac objects, but partially overlapping with the latter, consistent with the results of \cite{sun15}. In fact, once normalised by the mass of the central black holes, the jet powers are consistent with each other for these three types of AGNs, indicating the scalability of the jet. The authors argued that the central engine of RLNLS1s is apparently quite similar to that of blazars, despite the observational differences. Based on the detailed analysis of the $\gamma$-ray flux variability and spectral properties of five RLNLS1s, \cite{pal15} found that these $\gamma$-NLS1s are similar to FSRQs in terms of their average $\gamma$-ray photon index, however their $\gamma$-ray luminosities are lower than powerful FSRQs but higher than BL Lac objects. In view of the curvature in the $\gamma$-ray spectrum and flux variability amplitudes, the authors suggested that these $\gamma$-ray NLS1s could be similar to powerful FSRQs, but with low or moderate jet power. Although similarities have been found between RLNLS1s and blazars, we cannot yet perform a comparison between the host galaxies of blazars (commonly found in ellipticals) and the RLNLS1s of our sample. All our sources are at relatively high redshift, and their SDSS images are insufficient for host analysis. 

It has been suggested that outflows in AGNs can be accelerated by the radiation pressure of the accretion disk \cite[e.g.,][]{mur95}, which could be important for the jet production in sources
accreting at rates near the Eddington limit, as commonly found in NLS1 galaxies \cite[e.g.,][]{gru10,xu12}. However, \cite{cao14} found that it is difficult for the outflows to be directly driven from the photospheres of the radiation-dominated disks, which implies that hot gas (probably in the corona) is necessary for launching an outflow from the radiation-pressure-dominated disk. The authors investigated outflows accelerated from the hot corona above the disk by the magnetic field and radiation force of the accretion disk, and found that with the help of the radiation force, the mass-loss rate in the outflow is high, which however leads to a slow outflow. In our RLNLS1s sources, the accretion rate is high, and the radiation pressure could be significant. Therefore, the scenario of \cite{cao14} might be relevant in explaining why the jets in RLNLS1s are in general mildly relativistic compared with those in blazars. Recently, \cite{ber15b} proposed that the jet in RLNLS1s can be possibly launched by a rapidly accreting disk along with the winds when the angle between the poloidal component of the magnetic field and the accretion disk is less than a critical angle.

Alternatively, in the Blandford-Znajek (BZ) mechanism, the jet energy and angular momentum can be extracted from a rotating black hole \citep{bla77}. The low jet speed in RLNLS1s is consistent with the significant correlation between black hole mass and the jet bulk Lorentz factor found in typical blazars by \cite{cha12}, as the black hole masses of RLNLS1s are systematically lower than blazars. This correlation can be well explained in a black hole spin scenario, in the sense that the faster moving jets are magnetically accelerated by the magnetic fields threading the horizon of more rapidly rotating black holes \citep{cha12}. Recent investigation indeed suggested that most supermassive black holes in elliptical galaxies (e.g. for FSRQs) have on average higher spins than the black holes in spiral galaxies, where random, small accretion episodes (e.g., tidally disrupted stars, accretion of molecular clouds) might have played a more important role \citep{vol07}. If RLNLS1s do follow the correlation of \cite{cha12}, the low jet speed in RLNLS1s may then be due to the low spin. Few actual BH spin estimates for NLS1 galaxies are available to date, and none for the radio-loudest NLS1 galaxies. Based on deep X-ray observations of a few nearby (radio-quiet) NLS1 galaxies, intermediate to very high spins were estimated \cite[e.g.,][]{min09,fab12,fab13,par14,gal15}, while \cite{don13} preferred a low-spin solution, based on SED fitting of the NLS1 galaxy PG 1244+026. \cite{liu15} reported indications for low or intermediate average BH spin, estimated from stacked X-ray spectra of a larger sample of 51 NLS1 galaxies. Estimates for radio-loud NLS1 galaxies are not yet available.

Up to now, less than ten RLNLS1s were detected in $\gamma$-rays with $Fermi$/LAT \citep[see][]{abd09a,abd09b,dam12,dam15,fos11,yao15b}. The only source in our sample with a tentative $\gamma$-ray detection with $Fermi$, SDSS J124634.65+023809.0, has the steepest spectrum with a $\gamma$-ray photon index of $\Gamma=3.1$, however with the lowest Test Statistic (TS=15) \citep{fos11}. This source shows compact core-only morphology (see Fig. \ref{j1246}), different from the common core-jet structure in other $\gamma$-ray sources \cite[e.g.][]{ori12}. The brightness temperature of $10^{9.7}$ K is significantly lower compared to other $\gamma$-ray sources, which is usually larger than $10^{11}$ K \cite[e.g., PMN J0948+0022 in][]{gir11}. This is somewhat in contrast to the notion that the $\gamma$-ray objects tend to have larger beaming effects than non-$\gamma$-ray detected ones, for example, as suggested for BL Lac objects in \cite{wu14}. Finally, we would like to point out that the majority of the $\gamma$-ray detected NLS1 galaxies have published VLBI observations, and were therefore not re-analyzed in the study presented here. Therefore, our sample likely excludes a few of the most variable, and most highly beamed sources. We will present the results for a sample of fainter sources in a forthcoming paper based on the VLBA observations (Gu et al. 2015, in prep.). Moreover, in order to directly constrain the jet bulk speed, multi-epoch VLBA observations have been planned to detect the jet proper motion for a sample of RLNLS1s with resolved core-jet structure. All these observations will provide us with a deeper understanding on the jet formation in these low black hole mass, and high accretion rate systems.

\section{Conclusions}

As the first step of our systematic studies of the pc scale radio structure of RLNLS1s, we have obtained VLBA observations of 14 very radio-loud NLS1 galaxies in 2013. These observations reveal that the radio morphology of the sources is generally compact. Although all these NLS1 galaxies are very radio-loud objects with $R>100$, their jet properties are diverse, in terms of the mas scale morphology and overall radio spectral shape. The low core brightness temperatures, if confirmed with higher resolution, indicate that beaming effects are generally not significant, likely implying a low jet speed in these sources. This, in combination with the low kinetic/radio power, is likely responsible for the compact VLBA radio structure in most sources. The mildly relativistic jet in these high accretion rate systems could likely be explained by the scenario that jets/outflows are accelerated from the hot corona above the disk by the magnetic field and radiation force of the accretion disk, in which the jets/outflows are slow. Alternatively, the low jet bulk velocity can be explained by the Blandford-Znajek mechanism, if the black holes have low spin.



\acknowledgments

We thank the anonymous referee for constructive comments that improved the manuscript. 
We thank Xinwu Cao, Tao An and Thomas Krichbaum for valuable discussions. 
This work is supported by the National Science Foundation
of China (grants 11473054, 11273042, 11473035, 11173046, and U1531245) and by the Science and Technology Commission of Shanghai
Municipality (grants 14ZR1447100 and 12ZR1436100), the Strategic Priority Research Program ``The Emergence of Cosmological Structures'' of the Chinese Academy of Sciences (Grant No. XDB09000000), and the Strategic Priority Research Program on Space Science, the Chinese Academy of Sciences (Grant No. XDA04060700).
This study makes use of data from the SDSS (see
http://www.sdss.org/collaboration/credits.html).



{\it Facilities:} \facility{VLBA}

\clearpage



\clearpage









\clearpage

\clearpage

\begin{table}[tbh]
\caption{Sample of Radio-loud NLS1 Galaxies \label{source}} \vspace{2mm}
\begin{tabular}{lcclccccc}
  \hline\hline
Name          &    $z$    &   $f_{\rm 1.4GHz}$     &  $f_{\rm 5GHz}$ &   $\alpha$    &  log $R$ & a & LS  & r.s.    \\
                          &           &   (mJy)     & (mJy) &    &    &  (arcsec)    &  (kpc)    &            \\ 
(1)                       &   (2)     &   (3)       & (4)   &   (5)  &  (6)  & (7) & (8)   &   (9)      \\
\hline
SDSS J081432.11+560956.6    &   0.509   &   69/60       &   43 &  -0.33 &   2.53  &   2.84  &  ﻿17.4   &     F      \\ 
SDSS J085001.17+462600.5    &   0.523   &   21/16       &      &        &   2.23  &   1.26  &   7.8    &           S \\ 
SDSS J090227.16+044309.6    &   0.532   &   153/157    &  106 &  -0.30 &   3.02  &   1.00  &   6.3    &   F        \\  
SDSS J095317.09+283601.5    &   0.657   &   45/43       &      &        &   2.71  &   1.95  &  13.6    &          S$^a$ \\  
SDSS J103727.45+003635.6    &   0.595   &   27/28       &      &        &   2.66  &   1.16  &   7.7    &           F$^a$ \\  
SDSS J104732.68+472532.1    &   0.798   &   734/789    &  404 &  -0.51 &   3.87  &   1.16  &   8.7    &    S       \\  
SDSS J111005.03+365336.3    &   0.630   &   19/23       &      &        &   2.97  &   2.08  &  14.2    &           S \\  
SDSS J113824.54+365327.1    &   0.356   &   13/12       &      &        &   2.34  &   0.43  &   2.1    &            S  \\  
SDSS J124634.65+023809.0    &   0.362   &   37/36       &  46    &   0.17     &   2.38  &   1.12  &   5.6    &  F      \\  
SDSS J130522.75+511640.3    &   0.785   &   84/87       &  46  &  -0.50 &   2.34  &   1.34  &  10.0    &      S     \\  
SDSS J142114.05+282452.8    &   0.538   &   48.7/45.4         &  40  &  -0.15 &   2.31  &   1.28  &   8.1    &        F   \\  
SDSS J144318.56+472556.7    &   0.703   &   165/166     &   72 &  -0.67 &   3.07  &   1.18  &   8.4    &     S      \\  
SDSS J154817.92+351128.0    &   0.478   &   141/141     &  107 &  -0.22 &   2.84  &   0.51  &   3.0    &     F      \\  
SDSS J163401.94+480940.2    &   0.494   &   8/14         &      &        &   2.31  &   1.15  &   6.9    &             F \\  
\hline
\end{tabular}
\vspace*{1mm}\footnotesize\\Notes. --- Col. (1): source name; Col. (2) redshift; Col. (3): the FIRST and NVSS flux densities at 1.4 GHz (in the format of FIRST/NVSS); Col. (4): the flux density at 5 GHz from the literature; Col. (5): the spectral index between 1.4 and 5 GHz when available ($f_{\nu}\propto\nu^\alpha$); Col. (6): the radio loudness defined as $f_\nu \rm (1.4 GHz)$$/f_\nu \rm (4400{\AA})$; Col. (7): the major axis from FIRST; Col. (8): the upper limit on the source linear size calculated from the FIRST major axis; Col. (9): the radio spectrum based on low-resolution NED data (see Section \ref{ind.obj.}): F - flat spectrum, S - steep spectrum, $^a$ - uncertain (see text for details).
\end{table}



\clearpage

\begin{deluxetable}{lcccrccr}
\tablewidth{0pt}
\tablecaption{Results for the radio-loud NLS1 Galaxies \label{modelfit}}
\tablehead{
\colhead{Name}          &
\colhead{Comps.}         & \colhead{$S$}        & \colhead{$r$}  &
\colhead{$\theta$}          & \colhead{$a$}    &
\colhead{$b/a$}  &  \colhead{log $T_{\rm b}$}  \\
\colhead{}             &
\colhead{}      & \colhead{(mJy)}        & \colhead{(mas)}  &
\colhead{(deg)}          & \colhead{(mas)}    &
\colhead{}  &  \colhead{(K)} \\
\colhead{(1)}           & \colhead{(2)}      &
\colhead{(3)}          & \colhead{(4)}  &
\colhead{(5)}          & \colhead{(6)}    &
\colhead{(7)}  &  \colhead{(8)} }
\startdata

 SDSS J081432.11+560956.6   &    C     &   19.3   &    0.09   &    -89.4   &    0.41   &     1.00     &     10.1     \\
                            &          &    6.3   &    1.40   &     91.0   &    3.64   &     1.00     &              \\
 SDSS J085001.17+462600.5   &    C     &   11.0   &    0.08   &    -74.4   &    0.43   &     1.00     &      9.8     \\
 SDSS J090227.16+044309.6   &    C     &   49.7   &    0.26   &   -124.3   &    0.16   &     1.00     &     11.4     \\
                            &          &   22.9   &    0.62   &     53.3   &    0.58   &     1.00     &              \\
                            &          &   21.5   &    2.62   &     63.0   &    1.15   &     1.00     &              \\
                            &          &    7.5   &    4.13   &     54.9   &    0.73   &     1.00     &              \\
 SDSS J095317.09+283601.5   &    C     &   12.7   &    0.03   &     88.1   &    0.63   &     1.00     &      9.6     \\
                            &          &    1.4   &    2.08   &     32.7   &    0.68   &     1.00     &              \\
 SDSS J103727.45+003635.6   &    C     &   21.9   &    0.04   &   -156.7   &    0.59   &     0.25     &     10.5     \\
 SDSS J104732.68+472532.1   &    C     &   19.4   &    1.37   &     98.2   &    5.39   &     0.36     &      8.4     \\
 SDSS J111005.03+365336.3   &    C     &    8.8   &    0.13   &     -7.7   &    0.44   &     0.50     &     10.0     \\
 SDSS J113824.54+365327.1   &    C     &    9.5   &    0.06   &    -43.3   &    0.77   &     0.52     &      9.5     \\
 SDSS J124634.65+023809.0   &    C     &    8.7   &    0.13   &    176.1   &    0.62   &     0.46     &      9.7     \\
 SDSS J130522.75+511640.3   &    C     &   15.0   &    0.05   &     88.5   &    0.20   &     0.85     &     10.8     \\
                            &          &    8.9   &    5.97   &   -173.7   &    2.23   &     0.29     &              \\
 SDSS J142114.05+282452.8   &    C     &   27.1   &    0.10   &    -76.8   &    0.18   &     1.00     &     11.0     \\
                            &          &    5.8   &    0.66   &    118.4   &    0.44   &     1.00     &              \\
                            &          &    0.9   &    3.45   &    126.6   &    3.92   &     0.34     &              \\
 SDSS J144318.56+472556.7   &    C$^{a}$     &   26.1   &    0.45   &    -60.2   &    0.39   &     1.00     &     10.3     \\
                            &          &   18.0   &    0.99   &    124.4   &    0.93   &     1.00     &              \\
                            &          &    5.5   &    2.07   &    104.5   &    0.21   &     1.00     &              \\
                            &          &    2.7   &    3.43   &    102.9   &    1.47   &     1.00     &              \\
                            &          &    3.6   &    7.72   &     88.8   &    2.85   &     1.00     &              \\
                            &          &    6.4   &   11.19   &     99.9   &    2.81   &     1.00     &              \\
                            &          &    3.3   &   13.66   &    106.0   &    2.95   &     1.00     &              \\
 SDSS J154817.92+351128.0   &    C     &   14.1   &    0.04   &   -103.6   &    0.98   &     0.21     &      9.9     \\
                            &          &    4.5   &    7.05   &      5.6   &    1.65   &     0.27     &              \\
                            &          &    8.4   &   19.52   &     10.6   &    7.87   &     0.40     &              \\
                            &          &    6.2   &   59.30   &     26.2   &    6.84   &     0.78     &              \\
 SDSS J163401.94+480940.2   &    C     &    9.5   &    0.11   &   -152.9   &    0.29   &     1.00     &     10.1     \\

\enddata
\tablenotetext{-}{Col. (1): source name; Col. (2): components: C=core, $^{a}$ - core identification is uncertain (see text for details); Col. (3) flux density; Col. (4)-(5): component position, and its position angle; Col. (6): major axis; Col. (7): axial ratio; Col. (8): brightness temperature.}
\end{deluxetable}

\clearpage

\begin{landscape}
\begin{deluxetable}{lccrcrrcrrr}
\tablewidth{0pt}
\tablecaption{Archive data of Radio-loud NLS1 Galaxies \label{mfarchive}}
\tablehead{
\colhead{Name}  & \colhead{array}  & \colhead{obs. date}  & \colhead{$\nu$}       & \colhead{Comps.}          & \colhead{$S$}      &
 \colhead{$r$}  &
\colhead{$\theta$}          & \colhead{$a$}    &
\colhead{$b/a$}   & \colhead{log $T_{\rm b}$}  \\
\colhead{}  & \colhead{}  & \colhead{}  & \colhead{(GHz)}       & \colhead{}          & \colhead{(mJy)}      &
 \colhead{(mas)}  &
\colhead{(deg)}          & \colhead{(mas)}    &
\colhead{}   & \colhead{(K)}  \\
\colhead{(1)}  & \colhead{(2)}  & \colhead{(3)}  & \colhead{(4)}       & \colhead{(5)}          & \colhead{(6)}      &
 \colhead{(7)}  &
\colhead{(8)}          & \colhead{(9)}    &
\colhead{(10)}   & \colhead{(11)}  }
\startdata

SDSS J081432.11+560956.6   &  VLBA  &  2010-Mar-23   &   2.3   &  C  &   23.3  &   0.11   &   -76.0   &    0.14   &   1.00    &  11.8     \\
                           &        &                &   2.3   &     &    7.2  &   2.43   &    82.7   &    0.26   &   1.00    &           \\
                           &  VLBA  &  2006-May-31   &   5.0   &  C  &   30.7  &   0.09   &  -106.7   &    0.47   &   0.50    &  10.5     \\
                           &        &                &   5.0   &     &    8.5  &   1.85   &    93.6   &    1.55   &   0.32    &           \\
                           &  VLBA  &  2010-Mar-23   &   8.4   &  C  &   29.3  &   0.03   &   -88.0   &    0.06   &   1.00    &  11.5     \\
                           &        &                &   8.4   &     &    4.7  &   0.37   &    90.8   &    0.22   &   1.00    &           \\
                           &        &                &   8.4   &     &    1.9  &   2.55   &    89.5   &    0.55   &   1.00    &           \\
SDSS J104732.68+472532.1   &  VLBA  &  2006-May-01   &   5.0   &  C  &   21.9  &   0.31   &   -28.3   &    2.10   &   0.74    &  9.0      \\
                           &        &                &   5.0   &     &    9.5  &   4.05   &    85.0   &    1.74   &   0.22    &           \\
                           &  VLA   &  1990-Feb-20   &   8.4   &  C  &  144.8  &   7.90   &   143.5   &  114.49   &   0.47    &           \\
                           &        &                &   8.4   &     &   90.7  & 298.10   &   -41.5   &  357.23   &   0.63    &           \\
                           &        &                &   8.4   &     &   62.6  & 271.44   &   108.9   &   75.31   &   1.00    &           \\
                           &  VLA   &  2009-Sep-01   &  22.4   &     &  189.1  &  17.25   &   112.2   &  442.07   &   0.38    &           \\
SDSS J130522.75+511640.3   &  VLBA  &  2011-Mar-15   &   2.3   &  C  &   19.2  &   0.36   &   -10.3   &    0.27   &   1.00    &  11.2     \\
                           &        &                &   2.3   &     &    4.5  &   2.85   &  -146.9   &    1.24   &   1.00    &           \\
                           &        &                &   2.3   &     &    1.9  &   9.13   &  -158.1   &    0.79   &   1.00    &           \\
                           &  VLBA  &  2011-Mar-15   &   8.4   &  C  &   29.4  &   0.00   &   -22.6   &    0.09   &   1.00    &  11.3     \\
                           &        &                &   8.4   &     &    4.2  &   5.57   &  -172.5   &    1.24   &   0.35    &           \\
                           &        &                &   8.4   &     &    1.9  &   7.54   &  -169.9   &    1.73   &   0.72    &           \\
SDSS J142114.05+282452.8   &  VLBA  &  2010-Mar-23   &   2.3   &  C  &   33.0  &   0.04   &     5.4   &    0.67   &   1.00    &  10.6     \\
                           &        &                &   2.3   &     &    2.3  &   6.90   &   138.9   &    1.01   &   1.00    &           \\
                           &  VLBA  &  2010-Mar-23   &   8.4   &  C  &   21.6  &   0.08   &   136.2   &    0.48   &   0.29    &  10.1     \\

\enddata
\tablenotetext{-}{Col. (1): source name; Col. (2) radio telescope array; Col. (3) observational date; Col. (4): observing frequency; Col. (5): components: C=core; Col. (6): flux density; Col. (7)-(8): component position, and its position angle; Col. (9): major axis; Col. (10): axial ratio; Col. (11): brightness temperature.}
\end{deluxetable}
\end{landscape}





\begin{thebibliography}{}

\bibitem[Abdo et al.(2009a)]{abd09a} Abdo A. A. et al., 2009a, ApJ, 699, 976
\bibitem[Abdo et al.(2009b)]{abd09b} Abdo A. A. et al., 2009b, ApJ, 707, L142
\bibitem[An \& Baan(2012)]{an12} An, T., \& Baan, W.~A.\ 2012, \apj, 760, 77 
\bibitem[Angelakis et al.(2015)]{ang15} Angelakis, E., Fuhrmann, L., Marchili, N., et al.\ 2015, \aap, 575, A55
\bibitem[Becker et al.(1995)]{bec95} Becker R. H., White R. L., Helfand D. J., 1995, ApJ, 450, 559
\bibitem[Berton et al.(2015a)]{ber15a} Berton, M., Foschini, L., Ciroi, S., et al.\ 2015a, \aap, 578, A28
\bibitem[Berton et al.(2015b)]{ber15b} Berton, M., Foschini, L., Ciroi, S., et al.\ 2015b, arXiv:1506.05800 
\bibitem[Blandford \& Znajek(1977)]{bla77} Blandford, R.~D., \& Znajek, R.~L.\ 1977, \mnras, 179, 433
\bibitem[Blandford \& K\"{o}nigl(1979)]{bla79} Blandford R. D., K\"{o}nigl A., 1979, ApJ, 232, 34
\bibitem[Caccianiga et al.(2014)]{cac14} Caccianiga, A., Ant{\'o}n, S., Ballo, L., et al.\ 2014, \mnras, 441, 172 
\bibitem[Cao(2014)]{cao14} Cao, X.\ 2014, \apj, 783, 51 
\bibitem[Chai et al.(2012)]{cha12} Chai, B., Cao, X., \& Gu, M.\ 2012, \apj, 759, 114
\bibitem[Condon et al.(1998)]{con98} Condon, J.~J., Cotton, W.~D., Greisen, E.~W., et al.\ 1998, \aj, 115, 1693 
\bibitem[Cotton et al.(2003)]{cot03} Cotton, W.~D., Dallacasa, D., Fanti, C., et al.\ 2003, \pasa, 20, 12 
\bibitem[Dallacasa et al.(2013)]{dal13} Dallacasa, D., Orienti, M., Fanti, C., Fanti, R., \& Stanghellini, C.\ 2013, \mnras, 433, 147 
\bibitem[D'Ammando et al.(2012)]{dam12} D'Ammando, F., Orienti, M., Finke, J., et al.\ 2012, \mnras, 426, 317 
\bibitem[D'Ammando et al.(2013)]{dam13} D'Ammando, F., Orienti, M., Doi, A., et al.\ 2013, \mnras, 433, 952
\bibitem[D'Ammando et al.(2015)]{dam15} D'Ammando, F., Orienti, M., Larsson, J., \& Giroletti, M.\ 2015, \mnras, 452, 520
\bibitem[Doi et al.(2007)]{doi07} Doi A. et al., 2007, PASJ, 59, 703
\bibitem[Doi et al.(2011)]{doi11} Doi, A., Asada, K., \& Nagai, H.\ 2011, \apj, 738, 126 
\bibitem[Doi et al.(2012)]{doi12} Doi, A., Nagira, H., Kawakatu, N., et al.\ 2012, \apj, 760, 41
\bibitem[Done et al.(2013)]{don13} Done, C., Jin, C., Middleton, M., \& Ward, M.\ 2013, \mnras, 434, 1955 
\bibitem[Fabian et al.(2012)]{fab12} Fabian, A.~C., Zoghbi, A., Wilkins, D., et al.\ 2012, \mnras, 419, 116 
\bibitem[Fabian et al.(2013)]{fab13} Fabian, A.~C., Kara, E., Walton, D.~J., et al.\ 2013, \mnras, 429, 2917 
\bibitem[Fomalont(1999)]{fom99} Fomalont, E.~B.\ 1999, Synthesis Imaging in Radio Astronomy II, 180, 301 
\bibitem[Foschini(2011)]{fos11} Foschini, L.\ 2011, Narrow-Line Seyfert 1 Galaxies and their Place in the Universe, 24 
\bibitem[Foschini et al.(2015)]{fos15} Foschini, L., Berton, M., Caccianiga, A., et al.\ 2015, \aap, 575, A13 
\bibitem[Gallo et al.(2006)]{gal06} Gallo L. C. et al., 2006, MNRAS, 370, 245
\bibitem[Gallo et al.(2015)]{gal15} Gallo, L.~C., Wilkins, D.~R., Bonson, K., et al.\ 2015, \mnras, 446, 633 
\bibitem[Ghisellini et al.(1993)]{ghi93} Ghisellini, G., Padovani, P., Celotti, A., \& Maraschi, L.\ 1993, \apj, 407, 65
\bibitem[Giroletti et al.(2011)]{gir11} Giroletti, M., Paragi, Z., Bignall, H., et al.\ 2011, \aap, 528, LL11
\bibitem[Goodrich(1989)]{goo89} Goodrich R.W., 1989, ApJ, 342, 22
\bibitem[Gregory et al.(1996)]{gre96} Gregory P. C., Scott W. K., Douglas K., Condon J. J., 1996, ApJS, 103, 427
\bibitem[Grupe et al.(2010)]{gru10} Grupe, D., Komossa, S., Leighly, K.~M., \& Page, K.~L.\ 2010, \apjs, 187, 64 
\bibitem[Gu \& Chen(2010)]{gu10} Gu, M., \& Chen, Y.\ 2010, \aj, 139, 2612 
\bibitem[Ishwara-Chandra et al.(2010)]{ish10} Ishwara-Chandra, C.~H., Sirothia, S.~K., Wadadekar, Y., Pal, S., \& Windhorst, R.\ 2010, \mnras, 405, 436 
\bibitem[Kellermann \& Pauliny-Toth(1969)]{kel69} Kellermann K. I., Pauliny-Toth I. I. K., 1969, ApJ, 155, L71
\bibitem[Komossa et al.(2006)]{kom06} Komossa S., Voges W., Xu D.,  Mathur S., Adorf H. M., Lemson G., Duschl W., Grupe D., 2006, AJ, 132, 531
\bibitem[Komossa(2008)]{kom08} Komossa S., 2008, Rev. Mex. AA Ser. Conf., 32, 86
\bibitem[Komossa et al.(2015)]{kom15} Komossa, S., Xu, D., Fuhrmann, L., et al.\ 2015, \aap, 574, A121 
\bibitem[Kovalev et al.(2005)]{kov05} Kovalev, Y.~Y., Kellermann, K.~I., Lister, M.~L., et al.\ 2005, \aj, 130, 2473 
\bibitem[Kovalev et al.(2009)]{kov09} Kovalev, Y.~Y., Aller, H.~D., Aller, M.~F., et al.\ 2009, \apjl, 696, L17 
\bibitem[Kunert-Bajraszewska et al.(2006)]{kun06} Kunert-Bajraszewska, M., Marecki, A., \& Thomasson, P.\ 2006, \aap, 450, 945
\bibitem[Kunert-Bajraszewska \& Marecki(2007)]{kun07} Kunert-Bajraszewska, M., \& Marecki, A.\ 2007, \aap, 469, 437
\bibitem[Kunert-Bajraszewska et al.(2010)]{kun10} Kunert-Bajraszewska, M., Gawro{\'n}ski, M.~P., Labiano, A., \& Siemiginowska, A.\ 2010, \mnras, 408, 2261
\bibitem[Lane et al.(2014)]{lan14} Lane, W.~M., Cotton, W.~D., van Velzen, S., et al.\ 2014, \mnras, 440, 327 
\bibitem[Liu et al.(2015)]{liu15} Liu, Z., Yuan, W., Lu, Y., \& Zhou, X.\ 2015, \mnras, 447, 517 
\bibitem[Miniutti et al.(2009)]{min09} Miniutti, G., Panessa, F., de Rosa, A., et al.\ 2009, \mnras, 398, 255
\bibitem[Moran et al.(2000)]{mor00} Moran E. C., 2000, NewA Rev., 44, 527
\bibitem[Murray et al.(1995)]{mur95} Murray, N., Chiang, J., Grossman, S.~A., \& Voit, G.~M.\ 1995, \apj, 451, 498 
\bibitem[O'Dea(1998)]{ode98} O'Dea, C.~P.\ 1998, \pasp, 110, 493 
\bibitem[Orienti et al.(2012)]{ori12} Orienti, M., D'Ammando, F., Giroletti, M., \& for the Fermi-LAT Collaboration 2012, arXiv:1205.0402
\bibitem[Oshlack et al.(2001)]{osh01} Oshlack, A.~Y.~K.~N., Webster, R.~L., \& Whiting, M.~T.\ 2001, \apj, 558, 578
\bibitem[Osterbrock \& Pogge(1985)]{ost85} Osterbrock D. E., Pogge R. W., 1985, ApJ, 297, 166
\bibitem[Paliya et al.(2015)]{pal15} Paliya, V.~S., Stalin, C.~S., \& Ravikumar, C.~D.\ 2015, \aj, 149, 41
\bibitem[Parker et al.(2014)]{par14} Parker, M.~L., Wilkins, D.~R., Fabian, A.~C., et al.\ 2014, \mnras, 443, 1723 
\bibitem[Petrov(2013)]{pet13} Petrov, L.\ 2013, \aj, 146, 5 
\bibitem[Readhead(1994)]{rea94} Readhead A.C.S., 1994, ApJ, 426, 51
\bibitem[Rengelink et al.(1997)]{ren97} Rengelink, R.~B., Tang, Y., de Bruyn, A.~G., et al.\ 1997, \aaps, 124, 259 
\bibitem[Richards et al.(2015)]{ric15a} Richards, J.~L., Lister, M.~L., Savolainen, T., et al.\ 2015, IAU Symposium, 313, 139 
\bibitem[Richards \& Lister(2015)]{ric15} Richards, J.~L., \& Lister, M.~L.\ 2015, \apjl, 800, L8 
\bibitem[Spergel et al.(2003)]{spe03} Spergel, D.~N., Verde, L., Peiris, H.~V., et al.\ 2003, \apjs, 148, 175 
\bibitem[Sulentic et al.(2008)]{sul08} Sulentic, J.~W., Zamfir, S., Marziani, P., \& Dultzin, D.\ 2008, Revista Mexicana de Astronomia y Astrofisica Conference Series, 32, 51 
\bibitem[Sun et al.(2015)]{sun15} Sun, X.-N., Zhang, J., Lin, D.-B., et al.\ 2015, \apj, 798, 43 
\bibitem[Volonteri et al.(2007)]{vol07} Volonteri, M., Sikora, M., \& Lasota, J.-P.\ 2007, \apj, 667, 704 
\bibitem[Wajima et al.(2014)]{waj14} Wajima, K., Fujisawa, K., Hayashida, M., et al.\ 2014, \apj, 781, 75 
\bibitem[Wang et al.(2006)]{wan06} Wang, T.-G., Zhou, H.-Y., Wang, J.-X., Lu, Y.-J., \& Lu, Y.\ 2006, \apj, 645, 856 
\bibitem[Wu et al.(2013)]{wu13} Wu, F., An, T., Baan, W.~A., et al.\ 2013, \aap, 550, A113 
\bibitem[Wu et al.(2014)]{wu14} Wu, Z., Jiang, D., Gu, M., \& Chen, L.\ 2014, \aap, 562, A64 
\bibitem[Xu et al.(2012)]{xu12} Xu, D., Komossa, S., Zhou, H., et al.\ 2012, \aj, 143, 83
\bibitem[Yao et al.(2015a)]{yao15a} Yao, S., Yuan, W., Komossa, S., et al.\ 2015a, \aj, 150, 23
\bibitem[Yao et al.(2015b)]{yao15b} Yao S., Yuan W., Zhou H., et al. 2015b, \mnras, in press
\bibitem[Yuan et al.(2008)]{yua08} Yuan W., Zhou H. Y., Komossa S., Dong X. B., Wang T. G., Lu H. L., Bai J. M., 2008, ApJ, 685, 801
\bibitem[Zhou et al.(2006)]{zho06} Zhou H., Wang T., Yuan W., Lu H., Dong X., Wang J., Lu Y., 2006, ApJS, 166, 128
\bibitem[Zhou et al.(2007)]{zho07} Zhou H. et al., 2007, ApJ, 658, L13

\end{thebibliography}
\end{document}